\documentclass[a4paper,12pt]{article}

\usepackage{amsfonts,bm,amssymb,euscript,array,cite}
\usepackage{amsmath}
\usepackage[dvips]{epsfig}


\newcommand{\be}{\begin{equation}}
\newcommand{\ee}{\end{equation}}
\newcommand{\beq}{\begin{equation}}
\newcommand{\eeq}{\end{equation}}
\newcommand{\beqa}{\begin{eqnarray}}
\newcommand{\eeqa}{\end{eqnarray}}
\newcommand{\eq}[1]{(\ref{#1})}
\def\nn{\nonumber}
\def\bea{\begin{eqnarray}}
\def\eea{\end{eqnarray}}
\def\obar{\overline}
\def\a{\alpha}          
           
  \def\G{\Gamma}  
\def\d{\delta}  \def\D{\Delta}  
\def\e{\epsilon}

\def\l{\lambda} \def\L{\Lambda}

  \def\cC{{\cal C}}
\def\cD{{\cal D}}  \def\cF{{\cal F}}
 \def\cH{{\cal H}} 
  
 \def\cN{{\cal N}} \def\cO{{\cal O}}

\def\R{{\mathbb R}}
\def\C{{\mathbb C}}

\def\one{\mbox{1 \kern-.59em {\rm l}}}

\def\bit{\begin{itemize}}
\def\eit{\end{itemize}}
\def\Tr{\mbox{Tr}}

\def\({\left(}
\def\){\right)}
\def\tens{\otimes}

\def\reps{representations }

\allowdisplaybreaks[3]

\begin{document}

\begin{titlepage}
\begin{flushright}
LMU-TPS 05/01 \\
\end{flushright}

\vspace{.3cm}
\begin{center}
\renewcommand{\thefootnote}{\fnsymbol{footnote}}
{\Large \bf A non-perturbative approach to non-commutative \\
scalar field theory}
\vskip 20mm
{\large  {Harold Steinacker\footnote{harold.steinacker@physik.uni-muenchen.de}}}\\
\renewcommand{\thefootnote}{\arabic{footnote}}
\setcounter{footnote}{0}
\vskip 10mm
{\small
Department f\"ur Physik \\
Ludwig--Maximilians--Universit\"at M\"unchen \\
Theresienstr.\ 37, D-80333 M\"unchen, Germany}
\end{center}
\vspace{2cm}
\begin{center}
{\bf Abstract}
\end{center}
{\small

Non-commutative Euclidean scalar field theory is shown to have an 
eigenvalue sector which is dominated by a well-defined eigenvalue
density, and can be described by a matrix model.
This is 
established using regularizations  of $\R^{2n}_\theta$ via fuzzy
spaces for the free and weakly coupled case, and 
extends naturally to the non-perturbative domain. 
It allows to study the renormalization of the
effective potential using matrix model techniques, 
and is closely related to UV/IR mixing.
In particular we find a phase transition for the $\phi^4$ model
at strong coupling, to a phase which is identified with
the striped or matrix phase. 
The method is expected to be applicable in 4 dimensions, where
a critical line is found which terminates 
at a non-trivial point, with nonzero critical coupling. 
This provides evidence for a non-trivial fixed-point for
the 4-dimensional NC $\phi^4$ model.
}

\vspace{1cm}

\end{titlepage}

 \tableofcontents

\section{Introduction}

The idea of considering quantum field theory on ``quantized'' or
non-commutative spaces (NCFT) was put forward a long time ago 
\cite{heisenberg}, and
has been pursued vigorously in the past years; see e.g. 
\cite{douglas,szabo} for a review. 
A very intriguing phenomenon which was found 
in this context is the so-called UV/IR mixing \cite{seiberg}, 
linking the usual UV divergences to new singularities in the IR.
On a technical level, it arises because of a very different
behavior of planar and non-planar diagrams, which must be
distinguished on NC spaces. The planar diagrams 
are essentially the same as in the
commutative case. The non-planar diagrams however lead to oscillating
integrals, which are typically 
finite as long as the external momentum $p$ is
non-zero, but become divergent in the limit $p \to 0$. 
This leads to serious obstacles to perturbative 
renormalization \cite{seiberg}.
Furthermore, it appears to signal an additional phase
denoted as ``striped phase''  \cite{gubser,zappala,wu}, which arises
as the minimum of the effective action is no longer at zero momentum.

Because UV/IR mixing is so generic in the NC case, it is necessary
to come to terms with it and to find suitably adapted
quantization methods. The first step is clearly a suitable regularization
of the models. This can be achieved by parametrizing the fields
in terms of finite matrices, which is very natural on NC spaces.
Several such methods are available by now, using
e.g. using fuzzy spaces, non-commutative tori, etc.
The action for scalar fields 
is then a functional of a hermitean matrix $\phi$, where the
potential $Tr V(\phi)$ looks like a ``pure'' matrix model with 
$U(N)$ invariance, which is however broken by the kinetic term.
The UV/IR mixing is expected to be recovered in the
continuum limit. 

Such a regularization has been used recently to 
confirm numerically the non-trivial phase structure 
mentioned above in the non-commutative $\phi^4$ model \cite{bieten,martin}.
There has also been remarkable progress on the analytical side
using matrix techniques:
A modified $\phi^4$ model with an explicit IR regulator term in the action
was shown to be perturbatively renormalizable \cite{wulki}, 
and certain self-dual models of NC field theory 
were solved exactly using a matrix model formulation
\cite{szabo-langmann}.
For gauge theories, the
applicability of well-known techniques from random matrix theory
has also been shown in simple cases \cite{matrixsphere,szabo-gaugematrix}. 
For the  $\phi^4$ model, 
a similar approach using random matrix theory
does not seem possible at first sight, 
lacking $U(N)$ invariance. Nevertheless, it was
conjectured in \cite{martin} that the striped phase should be
identified with a ``matrix phase'' for the fuzzy sphere, where
the action appears to be dominated by a pure potential model in that phase.
Hence a simple analytical approach which allows 
to study also scalar NCFT's with
non-trivial phase structure and UV/IR mixing is highly desirable. 
In particular, it seems that
the obvious parallels between NCFT and pure matrix models due to
UV/IR mixing have not yet been fully exploited,
 apart from 
integrable cases \cite{szabo-langmann}.

The aim of this paper is to show that there is indeed
a simple matrix model
description which captures a certain sector of scalar NCFT, due to
UV/IR mixing. This suggests a new approach to scalar NCFT which
not only  provides new insights, but also 
new tools to study the renormalization
of the effective potential.
The starting point is an appropriate parametrization for the fields:
since  $\phi$ is a hermitian matrix, it can be
diagonalized as $\phi = U^{-1} diag(\phi_i) U$ where $\phi_i$ are the
real eigenvalues. Hence the field theory can be reformulated in terms
of the eigenvalues $\phi_i$ and the unitary matrix $U$. The main
observation of this paper
is now the following: the probability measure induced on
the (suitably rescaled) 
eigenvalues $\phi_i$ from the path integral is sharply localized, and 
described by an ordinary, simple matrix model.
This means that only fields $\phi$ with a particular eigenvalue
distribution characterized by a certain function $\rho(s): [-1,1] \to \R^+$
contribute significantly to the (euclidean) 
path integral. While this is plausible
using the above parametrization, it is a 
nontrivial statement which is only true in the con-commutative
regime. It is established first for the free case
(which therefore {\em does} know about non-commutativity, contrary to 
a common belief) and extends immediately to the interacting case 
at least on a perturbative level. This is directly related to UV/IR mixing,
since non-planar contributions to the eigenvalue
observables are suppressed by the oscillatory factors, while the
planar contributions can be described by a simple matrix model
without kinetic term. It is quite
obvious that this will extend also to the
non-perturbative level, in a suitable domain. This suggests that
scalar NCFT can be characterized by a single function $\rho(s)$.

We then work out some simple applications of this approach,
which do not require long computations. 
In the weak coupling regime, this leads to a very simple method of
computing the mass renormalization using matrix model techniques. 
In particular the standard one-loop result 
for the mass renormalization in the $\phi^4$ model
is recovered in a non-standard way,
and finds a natural interpretation in the matrix model. 
Unfortunately the running of the coupling does not seem
to admit such a simple computation. 
In any case, we will argue that there exists a scaling limit with 
a non-trivial correlation length in the
continuum limit, suggesting the existence of a renormalized
$\phi^4$ model in 2 and 4 dimensions. 

Extending these results to the non-perturbative regime, 
we find a phase transition for the $\phi^4$ model
in 2 and 4 dimensions, to a phase which is tentatively identified with
the striped or matrix phase of \cite{gubser,martin}. 
This can be compared with numerical results available for the fuzzy
sphere in 2 dimension, with reasonable agreement which
confirms the overall picture.
The method is expected to work better in 4 dimensions, 
due the stronger divergences which are crucial for our derivation. 
In the 4-dimensional case, the critical line is found to terminate 
at a non-trivial point, with nonzero critical coupling $g_c \neq 0$.
This is expected to be a sound prediction, suggesting 
the existence of a non-trivial fixed-point for
the 4-dimensional NC $\phi^4$ model, and hence renormalizability 
in an appropriate sense. 

To summarize, 
while the dominance of planar diagrams in NCFT is very well known, 
the description of the eigenvalue sector 
in terms of a simple matrix model \eq{tilde-S}
appears to be new and is very practical. 
This provides an alternative approach to some of the results of
\cite{gubser,zappala,wu} on the  phase structure in
NCFT, confirming the rough picture of a phase transition towards a
phase which breaks translational invariance. However, the phase
transition is predicted
 to be higher-order, as opposed to \cite{gubser,zappala,wu}.
We find in addition a critical coupling $g_c\neq 0$ in 4 dimensions, 
and it would be very interesting to verify
this numerically.

This paper is organized as follows. Section \ref{sec:setup} provides some
background recalling the UV/IR mixing, and introduces the matrix
regularizations used later. 
Section \ref{sec:EV} is the core of this paper: after identifying the
suitable observables, we show that the
eigenvalue distribution of free fields is given by Wigner's
semi-circle law, which allows to replace the kinetic term by the matrix
model \eq{eff-act-eig}. 
Interactions are included in Section \ref{subsec:interactions}.
The corresponding reformulation of scalar NCFT using
eigenvalue and angle coordinates is discussed in general in
Section \ref{sec:angle-eigenvalues} and \ref{subsec:physics},
including an intuitive semi-classical picture.
This is then applied to the $\phi^4$ model in section \ref{sec:phi4},
and related to some standard results for hermitean matrix models. 
These allow in Section \ref{sec:weak-coupling}
to obtain the mass renormalization in a very simple way. The phase
transition is studied in detail for 2 and 4 dimensions
in section \ref{sec:phase-transition},
and compared with numerical results for the fuzzy sphere.
We conclude in Section \ref{sec:discussion} 
with further remarks and an outlook.

\section{NC scalar fields and UV/IR mixing}
\label{sec:setup}

Consider scalar field theory on the non-commutative Moyal plane
$\R^d_\theta$ in even dimensions,
with action
\be
S = \int d^d x \(\frac 12 \partial_i \phi \partial_i \phi 
+ \frac 12 m^2 \phi^2 + \frac{g}{4} \phi\star\phi\star\phi\star\phi\).
\label{Rd-action-star}
\ee
Here $\phi$ is a function on $\R^d$, and 
the $\star-$ product is the standard Moyal product
of functions on $\R^d$, which can be
written as
\begin{align}
  (a\star b)(x) &= \int \frac{d^4k}{(2\pi)^4} \int d^4 y \; a(x{+}\tfrac{1}{2}
  \theta {\cdot} k)\, b(x{+}y)\, \mathrm{e}^{\mathrm{i} k \cdot y}\;,
\label{starprod}
\\*
& (\theta {\cdot} k)_i = \theta_{i j} k_j\;,\quad k{\cdot}y = k_i
y_i\;,\quad \theta_{i j}=-\theta_{j i}\;.  \nonumber
\end{align}
This can be understood as a pull-back of the operator product of 2
operators $a$ and $b$ from a representation of the underlying
(Heisenberg) algebra
\be
[x_i,x_j] = i\theta_{ij},
\label{theta-def}
\ee
using a suitable quantization map. 
We assume that $\theta_{ij}$ is non-degenerate in this paper.

The model \eq{Rd-action-star} 
written down above is not well-defined as it stands, and
needs regularization. The simplest way to proceed is to use a
sharp UV cut-off $\L$, which leads to standard computations
and will be justified below.
The perturbative quantization of \eq{Rd-action-star} 
differs from the commutative case by
the fact that planar and non-planar diagrams must be distinguished. 
The reason is that commuting 2 plane waves with wavenumbers $k$ and 
$k'$ produces a factor $e^{-i k \theta k'}$, which makes non-planar 
loops convergent {\em for generic external momenta}. More explicitly, 
the basic one-loop planar and non-planar self-energy diagrams (without
counting symmetry factors) are \cite{seiberg}
\bea 
\G^{(2)}_P := \int \frac{d^dk}{(2\pi)^d}\; \frac 1{k^2+m^2},\nn\\
\G^{(2)}_{NP}(p) := \int \frac{d^dk}{(2\pi)^d}\; 
   \frac{e^{i k\theta p}}{k^2+m^2}.
\label{G-oneloop}
\eea 
$\G^{(2)}_{NP}(p)$ is finite as long as $p\neq 0$ due to the
oscillating term, but has an IR
singularity as $p \to 0$ because the $k$-integral is then divergent as
 usual. This is known as UV/IR mixing \cite{seiberg}, 
and appears to be a central feature of NC field theories.
It is a serious obstacle to perturbative renormalization, 
which was only overcome recently in a modified $\phi^4$ 
model \cite{wulki}.

In this paper, we shall try to turn this UV/IR mixing into a virtue, and 
point out that it is closely related to
an interesting property of the scalar field $\phi$ in the operator formulation,
which seems very useful and does 
not hold for ordinary field theories: The dynamical field 
$\phi$ has a well-defined eigenvalue distribution upon quantization,
which is governed
by a simple matrix model and can be studied using a saddle-point
analysis.

\subsection{Matrix regularization of $\R^n_\theta$}
\label{subsec:regularization}

Recall that if the non-commutative algebra \eq{theta-def} is represented 
on a (infinite-dimensional) Hilbert space $\cH$, the integral is 
given by the suitably normalized trace:
\be
\int d^d x f(x) = (2\pi)^{d/2} \sqrt{\det\theta}\; Tr f 
= (2\pi \theta)^{d/2}\; Tr f 
\ee
where $f$ in the rhs is the operator version of $f(x)$
(as obtained e.g. using the Weyl quantization map). 
This is a manifestation of the Bohr-Sommerfeld quantization condition,
relating the volume of the phase space to the dimension of the Hilbert
space. 
The last line holds for $Sp(d)$- invariant $\theta_{ij}$, 
which we assume in this paper for simplicity. 

We want to approximate this using some {\em
finite-dimensional} matrix algebra. This can be achieved e.g.
using a suitable scaling limit of the fuzzy sphere for $d=2$, or more generally
fuzzy $\C P^n$ for $d=2n$, see Appendix A.
Indeed fuzzy $\C P^n \to \R^{2n}_\theta$ in a suitable limit, where
$\theta_{ij}$ turns out to be invariant under $Sp(2n)$, and the
propagator is the usual one with a sharp momentum cutoff. 
Another possible regularization is using so-called
NC lattices which are products of certain (``fuzzy'') NC  tori
\cite{ambjorn}, see also Appendix A. 
These are technically somewhat easier to handle, but lead to a modified
behavior of the propagators for large momenta; then the most 
general $\theta_{ij}$ can be obtained.
In all these regularizations, $\phi$ is a hermitian $\cN \times \cN$
matrix in some finite matrix algebra $Mat(\cN,\C)$, and  the model 
\eq{Rd-action-star} is replaced 
by a matrix model (where the kinetic term breaks the $U(\cN)$ symmetry).
In particular, the trace is now over the $\cN$ -dimensional
Hilbert-space $\cH = \C^\cN$, where $\cN$ is related to 
the cutoff $\L$ and $\theta$ in a specific way
\eq{cutoff-CPn}, \eq{NC-lattice-cutoff} depending on the regularization.
Then $V := \int d^d x 1 = (2\pi \theta)^{d/2}\; \cN$, and
we can write 
\be
\frac 1V\int d^d x f(x) = \frac 1{\cN}\; Tr f.
\label{integral-trace}
\ee
In particular, integrals of the type $\int \phi^{2n}$
depend only on the eigenvalues of $\phi$, and these are the
observables we want to study.

To make the paper most readable, the 
regularization using fuzzy $\C P^n$ with
sharp momentum cutoff $\L$ will be understood, while
using the conventional language of $\R^d_\theta$ as much as possible.
The results for general (non-degenerate) $\theta_{ij}$ 
and somewhat modified propagators would be qualitatively the same.

\section{The eigenvalue distribution of the scalar field} 
\label{sec:EV}

The basic idea is the following: Having regularized the 
model \eq{Rd-action-star}
in terms of a finite-dimensional hermitean matrix $\phi$, we can 
diagonalize it as
\be
\phi = U^{-1} (\phi_i) U
\ee
where $U$ is a unitary $\cN \times \cN$ matrix, and 
$(\phi_i) \equiv diag(\phi_1, ...,\phi_\cN)$ is diagonal 
with real eigenvalues. 
The integration measure in the path integral can now be written as 
$\int \cD \phi e^{-S}= \int d\phi_i \D^2(\phi_i) \int dU e^{-S}$,
where $\D^2(\phi_i)  =\prod_{i<j} (\phi_i-\phi_j)^2$ is the
Vandermonde-determinant and $dU$ the Haar measure for $SU(\cN)$.
We are interested in the 
probability measure or effective action for these eigenvalues, induced
by this path integral. For this purpose,
consider e.g. the expectation values
\be
\langle \int d^d x \phi^{2n}(x) \rangle 
= \frac 1Z \int \cD \phi 
\exp(-S) (\int d^d x \phi^{2n}(x)).
\label{intphi2n-def}
\ee
They are strongly divergent normally but make sense in the regularized
(matrix) case; in particular, we do {\em not} want to replace 
the $\phi^{2n}(x)$ by
some renormalized objects but simply keep track of their dependence on the
cutoff. Since they depend only on the 
eigenvalues of the field $\phi$ in the matrix representation, 
we can determine the effective eigenvalue
distribution by studying such observables. This turns out to be
non-trivial already in the free case:

\subsection{The free case}
\label{subsec:free distri}

We compute the observables \eq{intphi2n-def} for
$g=0$  with a sharp UV - cutoff $\L$ using Wicks theorem. 
This involves in general planar and
non-planar diagrams. The simplest case is\footnote{we basically assume
  that $\theta_{ij}$ is non-degenerate in this paper, which implies
  that $d$ is even. However, some of the results  extend to the 
degenerate case, which will be pursued elsewhere}
\bea
\langle \int d^d x \phi^{2}(x) \rangle 
&=& V\;\int_0^\L \frac{d^d p}{(2\pi)^d}\; \frac 1{p^2+m^2}
= \frac{V}{2^{d-1}\pi^{d/2}(d/2-1)!}\int_0^\L dp \frac{p^{d-1}}{p^2+m^2}\nn\\
&=:& c(m,\L) V\;\L^{d-2}.
\label{phi2-expect}
\eea
This formal result will be fully justified in Appendix A
using regularizations of $\R^{d=2n}_\theta$ in terms of 
fuzzy $\C P^n$ \eq{CPn-comput} or fuzzy tori. Here
$V$ denotes the regularized volume of $\R^d_\theta$,
and $c = c(m,\L)$ is of order 1 for $d \geq
3$, and $c =O(\ln\L)$ for $d=2$. More precisely, in 4 dimensions we have
\bea
\langle \int d^4 \phi^{2}(x) \rangle 
&=& \frac{V m^{2}}{8\pi^{2}}\int_0^{\L/m} du \frac{u^{3}}{u^2+1}  
= \frac{V \L^2}{16\pi^{2}}\;
\(1 - \frac{m^2}{\L^2}\ln(1 + (\frac{\L}{m})^2)\) 
\label{phi2-expect-4d}
\eea
and in 2 dimensions
\be
\langle \int d^2 \phi^{2}(x) \rangle 
= \frac{V}{2\pi}\int_0^{\L/m} du \frac{u}{u^2+1}  
= \frac{V}{4\pi}\ln(1+\frac{\L^2}{m^2})
\label{phi2-expect-2d}
\ee
(the subleading behavior is modified for the regularization using NC
tori, see Appendix A).
Therefore
\be
c(m,\L) = \left\{\begin{array}{ll} \frac{1}{16\pi^{2}}\;
\(1 -\frac{m^2}{\L^2}\ln(1+\frac{\L^2}{m^2})\), & d=4\\
                 \frac{1}{4\pi}\ln(1+\frac{\L^2}{m^2}))  , & d=2.
                \end{array}\right.
\ee
Next, consider
\bea
\langle \int d^d x \phi(x)^{4} \rangle &=& 
\langle \int d^d x \phi(x)^{4} \rangle_{Planar} 
   + \langle \int d^d x \phi(x)^{4} \rangle_{Non-Planar} \nn\\
&=& 2 V\;\int_0^\L \frac{d^d p}{(2\pi)^d}\; \frac 1{p^2+m^2}\;\G^{(2)}_P(p) 
  +  V\; \int_0^\L \frac{d^d p}{(2\pi)^d}\;\frac 1{p^2+m^2}\;\G^{(2)}_{NP}(p),\nn
\eea 
which is obtained by summing over all complete contractions of a vertex
with 4 legs. There are 2 planar and one non-planar such contractions,
the latter being given by joining the external legs of the 
non-planar self-energy diagram \eq{G-oneloop}. 
The planar contribution is simply
\be
\langle \int d^d x \phi(x)^{4} \rangle_{Planar}
 = 2 V\;c^2\L^{2(d-2)},
\ee
since $\G^{(2)}_P(p)$ is independent of $p$. 
On the other hand, the non-planar contribution
is subleading (this will be discussed in detail below), 
since $\G^{(2)}_{NP}(p)$ is finite except for the 
singularity at $p\to 0$. This is clearly related to UV/IR  mixing,
even though we are considering only the free case up to now.
We therefore expect
\be
\frac{\langle \frac 1V\int d^d x \phi(x)^{4} \rangle}
  {\langle \frac 1V\int d^d x \phi(x)^{2} \rangle^2} 
= 2\;\; + O(\frac1{c\L^{d-2}}).
\label{planar-approx}
\ee
More generally, consider
$$
\frac{\langle \frac 1V\int d^d x\phi(x)^{2n}\rangle}
{\langle  \frac 1V\int d^d x \phi(x)^{2} \rangle^n} 
=  \frac{\langle  \frac 1V\int d^d x \phi(x)^{2n} \rangle_{Planar}}
 {\langle  \frac 1V\int d^d x \phi(x)^{2} \rangle^n}
 +  \frac{\langle  \frac 1V\int d^d x \phi(x)^{2n} \rangle_{Non-Planar}}
 {\langle  \frac 1V\int d^d x \phi(x)^{2} \rangle^n}.
$$
The first term is of order one, and simply counts the number 
$N_{Planar}(2n)$ of planar contractions of a
vertex with $2n$ legs. 
The non-planar contributions always involve oscillatory
integrals, and do not contribute to the above ratio in the large $\L$ limit. 
Let us discuss them in more detail:
Assume that the cut-off is much larger
than the NC scale,
\be
\L^2 \theta \gg 1,
\ee
which will be understood from now on.
By rescaling the momenta $k' = k/\L$, the above ratios for $d \geq 3$ 
have the form
\be
R_{NP} :=\frac{\langle  \frac 1V\int d^d x \phi(x)^{2n} \rangle_{Non-Planar}}
{\langle  \frac 1V\int d^d x \phi(x)^{2} \rangle^n}
 \approx \int_0^1 \frac{d^dk'_{1}}{(k'_1)^{2}} ... \frac{d^d k'_n}{(k'_n)^{2}}
e^{i \L^2\sum k'_i \Theta k'_j} \quad\to 0\;\;\mbox{as}\;\;\L\to\;\;\infty,
\label{RNP-lemma}
\ee
which vanish for large $\L$ 
due to the rapidly oscillating exponential.
The integrals are over the unit ball resp. hypercube. 
This is established more carefully in Appendix B, where we show that 
$R_{NP} = O(1/\L)$ (at least) for 
$d \geq 3$. In 2 dimensions, these considerations are more delicate 
as the above ratio $R_{NP}$ vanishes only logarithmically, and
divergences may also arise in the IR.
One must then assume furthermore that e.g. $\frac{m}{m_\theta}$
is fixed while $\L \to \infty$, where
\be
m_\theta^2 := \frac 1{\theta}
\label{m-theta}
\ee
is the non-commutative mass scale.
Due to this weaker logarithmic behavior, 
the fluctuations are expected to be larger for $d=2$ than for $d=4$. 
We refrain here from more precise estimates which should be made in
the context of the regularized models, see Appendix B.

With all these assumptions, we conclude that
\be
\frac{\langle  \frac 1V\int d^d x \phi(x)^{2n} \rangle}
{\langle  \frac 1V\int d^d x \phi(x)^{2} \rangle^n}
= \frac{\langle  \frac 1V\int d^d x \phi(x)^{2n} \rangle_{Planar}}
 {\langle  \frac 1V\int d^d x \phi(x)^{2} \rangle^n}
= N_{Planar}(2n) 
\label{phi2n-exp}
\ee
in the large $\L$ limit. 
Notice that this result is very different from the conventional field
theory: the above calculation with a naive cutoff would have the same 
contributions from planar and non-planar diagrams. Then the 
total number of contractions of e.g. 
$\langle \phi^{2n}\rangle$ is of order $2^n n! \gg N_{Planar}(2n)$, 
which would invalidate the conclusions below.

Next, consider expectation values of products 
$\langle(\int d^d x \phi(x)^{2n_1})...(\int d^d x \phi(x)^{2n_k}) \rangle$.
We claim that this factorizes in the large $\L$ limit,
\bea
\frac{\langle\big( \frac 1V\int d^d x \phi(x)^{2n_1}\big)...
\big( \frac 1V\int d^d x \phi(x)^{2n_k}\big) \rangle}
{\langle  \frac 1V\int d^d x \phi(x)^{2} \rangle^{n_1}...
\langle  \frac 1V\int d^d x \phi(x)^{2} \rangle^{n_k}}
&=& \frac{\langle \frac 1V\int d^d x \phi(x)^{2n_1}\rangle}
  {\langle  \frac 1V\int d^d x \phi(x)^{2} \rangle^{n_1}}\; ...
\;\frac{\langle \frac 1V\int d^d x \phi(x)^{2n_k} \rangle}
{\langle  \frac 1V\int d^d x \phi(x)^{2} \rangle^{n_k}} \nn\\
&=& N_{Planar}(2n_1)\; ... \;N_{Planar}(2n_k)
\label{cluster}
\eea
This can be seen again in terms of contractions, because
propagators joining different vertices must satisfy additional
momentum constraints as opposed to the disjoint contractions,
therefore only the disjoint contributions survive in the large $\L$
limit. This amounts essentially to the cluster property.

Now recall that in the non-commutative case, the integral is 
given by the suitably normalized trace \eq{integral-trace}.
Hence the integrals above depend only on the eigenvalues of $\phi$, 
and we obtain statements about the induced eigenvalue
distribution. The
above observables \eq{cluster} can be written in the form
\be
(\frac 1{\cN} Tr \phi^{n_1}) ... (\frac 1{\cN} Tr \phi^{n_k}),
\label{Trphi-exp}
\ee
and completely determine the effective 
probability measure for the eigenvalues of $\phi$.
We can certainly write any such expectation value as
\be
\langle (\frac 1{\cN}  Tr \phi^{n_1}) ... 
(\frac 1{\cN}  Tr \phi^{n_k})\rangle = 
\int d \phi_1 ... d\phi_n \mu(\phi_1, ... ,\phi_n)\;
(\frac 1{\cN}  \sum_i \phi_i^{n_1}) ... (\frac 1{\cN}  \sum_j \phi_j^{n_k})
\label{mu-measure}
\ee
for some measure $\mu(\phi_1, ... ,\phi_n)$, which we would like
to determine for large $\L$.
In order to absorb the infinities we 
introduce a scaling constant 
\be
\a_0^2(m) = 4c\L^{d-2}
 = \left\{\begin{array}{ll} \frac{1}{4\pi^{2}}\;
\L^2 \(1 - \frac{m^2}{\L^2}\ln(1+\frac{\L^2}{m^2})\), & d=4\\
                 \frac{1}{\pi}\ln(1+\frac{\L^2}{m^2})  , & d=2
                \end{array}\right.
\label{alpha}
\ee 
and write\footnote{$\a_0$ is {\em not} the
wavefunction renormalization}
\be
\phi = \a_0 \varphi.
\label{varphi-scale}
\ee
Then \eq{phi2-expect} gives
\be
\langle \frac 1{\cN} Tr \varphi^{2} \rangle = \frac 14. 
\ee
Now all expectation values 
$\langle\frac 1{\cN} Tr \varphi^{2n}\rangle$
are finite and have a well-defined
limit $\cN \to\infty$. We can then describe the
eigenvalues of $\varphi$ (in increasing order) by an
eigenvalue distribution,
\be
\varphi(s) = \varphi_j, \qquad   s = \frac j{\cN},  
\qquad s \in [0,1].
\label{varphi-function}
\ee
Then e.g. 
\be
\frac 1{\cN} \Tr f(\varphi) 
= \frac 1{\cN} \sum_i f(\varphi_i) \to \int_{0}^1 ds\; f(\varphi(s))
\ee
in the large $\cN$ limit. The measure $\mu$ now becomes a measure 
$\mu[\varphi(s)]$ on 
the space of (increasing) functions $\varphi(s):[0,1] \mapsto \R$.
To find this measure $\mu[\varphi(s)]$, 
we first note that the factorization
property \eq{cluster} implies that the measure $\mu$ is {\em localized}, 
i.e. 
\be
\langle f(\varphi_i) \rangle = f(\obar\varphi(s))
\ee
for any function $f$,
where $\obar\varphi(s)$ is the (sharp and dominant) 
saddle-point or maximum of $\mu[\varphi]$.
This saddle-point $\obar\varphi(s)$
corresponds to a density of eigenvalues 
\be
\rho(\obar\varphi) = \frac{ds}{d\obar\varphi}, \qquad
\int_{-\infty}^{\infty} \rho(p) dp =1.
\label{density-phi}
\ee
The expectation value of the above observables is then given by
\be
\langle \frac 1{\cN} \Tr f(\varphi) \rangle
=  \int dp \rho(p)\; f(p),
\label{obs-expect-measure}
\ee
or $\langle \frac 1{\cN} \Tr \varphi^n \rangle
=  \int_{0}^1 ds\; \obar\varphi^n$; for  example,
$\langle\int ds \varphi(s)^{2} \rangle 
= \int ds \obar\varphi(s)^{2} 
= \frac 14$.
We want to determine this saddle-point $\obar\varphi(s)$.
This can be extracted from the above results: \eq{phi2n-exp} implies that
\be
\langle \int_{0}^1 ds \varphi(s)^{2n} \rangle 
= \int  \rho(p) p^{2n} dp
= (\frac 14)^n\; N_{Planar}(2n). 
\label{exp-2n-wigner}
\ee
There is a unique eigenvalue distribution with these properties,
given by the famous Wigner semi-circle law 
\be
\rho(p) = \left\{\begin{array}{ll}\frac 2{\pi} \sqrt{1-p^2} &  \;\; p^2<1\\
                           0, & \mbox{otherwise}.
                \end{array}\right.
\label{wigner-law}
\ee
This means that the eigenvalues of $\phi$ are distributed
correspondingly in the interval 
\be
\phi_i \in [-\a_0,\a_0].
\ee
It is fun to verify \eq{exp-2n-wigner} explicitly for small n;
indeed
\be
\int_{-1}^1 \rho(x) x^{2n} dx = \frac{\Gamma(\frac 12+n)}{\Gamma(2+n)}
= \frac 1{4^n}\; N_{Planar}(2n). 
\ee
In general,
this follows also from the basic properties of matrix models:
Consider the Gaussian matrix model with action
\be \fbox{$
\tilde S_0  = f_0(m) + \frac{2\cN}{\a_0^2} Tr \phi^2  $}
\label{eff-act-eig}
\ee
where $\phi = \a_0 \varphi$ 
is a $\cN \times \cN$ matrix. Here
$f_0(m)$ is some numerical function of $m$ which will be determined below, 
and $\a_0$ depends on $m$ via $c = c(m)$.
$\tilde S_0$ will find another interpretation 
in Section \ref{sec:angle-eigenvalues}.
$\tilde S_0$ is known to reproduce precisely the eigenvalue
distribution \eq{wigner-law} in the large $\cN$ limit \cite{BIPZ},  
and \eq{exp-2n-wigner} follows 
because again only planar diagrams contribute.
Indeed, one finds again e.g. 
\be
\langle \frac 1\cN Tr\phi^2\rangle 
= \a_0^2 \langle\frac 1\cN  Tr \varphi^2 \rangle = \frac 14\; \a_0^2 
\ee
etc.; we refer to the vast literature available on this subject, 
e.g. \cite{BIPZ,ginsparg,eynard}.

We conclude that if one is only interested in the eigenvalues, 
the free action 
$S_0 = \int d^d x \frac{1}{2}(\partial_i\phi \partial_i \phi + m^2 \phi^2)$ 
can be replaced by the effective action 
$\tilde S_0$ in
\eq{eff-act-eig}. Moreover, one can determine 
$f_0(m)$ such that 
the partition function is also recovered as
\be
Z  = \int \cD\phi e^{-S_0}
  = \int \cD\phi e^{-\tilde S_0} = 
 \int \cD \phi \exp(-f_0(m) -\frac{2\cN}{\a_0^2} Tr \phi^2);
\label{eff-act-eig2}
\ee
this will be understood better in Section \ref{sec:angle-eigenvalues}.
To summarize, we noted that the observables \eq{cluster} depend only on the
eigenvalues; the factorization property implies that
the dominating contribution comes from a well-defined eigenvalue
distribution, which is via \eq{exp-2n-wigner} 
identified as the Wigner-distribution
corresponding to a Gaussian matrix model. We can then write down an
effective Gaussian matrix model which reproduces all the expectation
values for these observables.  Note that 
the details of the propagator enter only through $\a_0$,
and the basic result depends only on the degree of divergence. 

It is interesting to compare this with the commutative case,
where the eigenvalues are replaced by the values of the field
$\phi(x)$ at each point, i.e. $N^d$ variables for a lattice regularization.
Using again a cutoff $\L$, we would have 
\be
\frac{\langle  \frac 1V\int d^d x \phi(x)^{2n} \rangle}
{\langle  \frac 1V\int d^d x \phi(x)^{2} \rangle^n}
= N_{Planar}(2n) + N_{Nonplanar}(2n)
 = \int_{-\infty}^{\infty} e^{-s^2/2} s^{2n}
\approx 2^n n! \gg N_{Planar}(2n).\nn
\ee
This could be reproduced by a Gaussian distribution
$\tilde S_0^{C}  = \frac{1}{2\a_0^2} \int d^d x \phi(x)^2$.  
The crucial difference is that this would describe $N^d$ 
independent variables (which is false of course, 
but ok for these observables) with a very
flat distribution, while in the NC case we consider only $\cN = N^{d/2}$
eigenvalues which are 
collective and intrinsically non-local variables, governed by 
an effective 
action with a sharp potential\footnote{recall that the 
saddle-point approximation for the
  matrix models is good for the eigenvalues, but not for the matrix
  elements};
note also the explicit factor $\cN$ in \eq{eff-act-eig}.
This will  become  more explicit in the following sections, which
do not apply to the commutative case.

Let us try to interpret this result. It may be
surprising that already the free scalar NCFT is apparently very
different from the commutative case, even though they are supposed to
be the same from the star-product point of view.
The reason is that we are looking at statistical properties of the
operator representation of the wavefunctions, which is related to their 
point-wise values only for low momenta where
the functions are ``almost-commutative''.  Therefore the non-classical 
properties of the high-energy modes are responsible for this property.
This already indicates
UV/IR mixing: the modes with $k \approx 0$ are suppressed 
upon quantization because they
tend to have the ``wrong'' eigenvalue distribution (in particular
$\phi_{k=0} \propto \one$ has only one  eigenvalue).

The existence of a well-defined eigenvalue distribution 
and its matrix model description will generalize
easily to the interacting case at least on a perturbative level, and
is very plausible also in the non-perturbative domain. 
Furthermore it suggests new and practical insights to the coupling
constants and their renormalization. This will be discussed next.

\subsection{Interactions}
\label{subsec:interactions}

If we include interactions of the form 
$$
S_{int}(\phi) = \frac{g_n}{n}\int d^d x \phi^n(x) 
= \frac{g_n}{n} (2\pi\theta)^{d/2}
 Tr \phi^n,
$$
the results \eq{eff-act-eig}, \eq{eff-act-eig2} 
for the free case generalize at least
on a perturbative level: 
consider
\be
Z_{int} = \int \cD\phi e^{-(S_0+S_{int})}
= \int \cD\phi e^{-S_0}\(1-\frac{g_n}{n} (2\pi\theta)^{d/2}
 Tr \phi^n + ...\)
\label{Z-perturb}
\ee
Hence the expansion of $e^{-S_{int}}$ 
in powers of $g_n$ leads
to additional terms of the form \eq{cluster} or \eq{Trphi-exp}, 
which are again evaluated by the rule \eq{obs-expect-measure}
for the  measure $\rho$ 
of the free case\footnote{This is valid as long as the coupling constants
$g_n$ are small enough so that the eigenvalue 
distribution remains unchanged.
This should be enough to determine the perturbation series, where 
$g_n$ can be arbitrarily small. In particular, it suffices to
determine mass renormalizations etc. as seen below.}.
We can write a similar formula to evaluate
observables such as $\langle \frac 1{\cN} Tr \phi^{2n} \rangle_g$ in the
presence of $g_n$.
Therefore the results of Section \ref{subsec:free distri} apply, 
and one can simply replace $S_0$ by $\tilde S_0$ in \eq{Z-perturb}.
We conclude that at least perturbatively, 
the eigenvalue sector of the NCFT with action
\be
S =  \int d^d x \frac{1}{2}(\partial_i\phi \partial_i \phi + m^2 \phi^2) +
    S_{int}
\ee
for any polynomial interaction is described by the {\em effective matrix
model}
\be \fbox{$
\tilde S(\phi) = f_0(m) +\frac{2\cN}{\a_0^2(m)} Tr \phi^2 + S_{int}(\phi) $}
\label{tilde-S}
\ee
where $\a_0 =\a_0(m)$ is given by \eq{alpha}, in the large $\cN$
limit. This is expected to be correct as long as the eigenvalue
distribution is close enough to the free one \eq{wigner-law}.
The partition function is similarly given by 
\be
Z_{int} =  \int \cD \phi \exp(-f_0(m) -\frac{2\cN}{\a_0^2(m)} Tr
\phi^2 + S_{int}(\phi)).
\ee
This defines by the usual matrix model technology an analytic function
in the couplings $g_n$.
In the later sections, this will allow us in particular to determine some 
renormalization properties of the potential is a very simple way.
We will also explore non-perturbative implications such as 
phase-transitions, hoping that this will give at least qualitatively
correct results.

\paragraph{Scaling and relevant couplings.}

Let us try to estimate the impact of the interaction terms to the
eigenvalue distribution. In the matrix-regularizations of NCFT used
below, we will have
$\cN \sim N^{d/2}$ and $\L = O(\sqrt{N})$. Therefore the following
scaling behavior holds
\be
\phi \sim \a_0 \sim  \L^{\frac d2 -1},\quad \cN = \L^d
\ee
using \eq{varphi-scale}. We assume this scaling also 
in the interacting case in the weakly-coupled domain,
and check for which $g_n$ this is self-consistent. 
Then \eq{tilde-S} essentially becomes 
$\tilde S \sim Tr(\cN \varphi^2 + g_n \a_0^n \varphi^n) 
=: Tr(\cN V(\varphi))$, 
omitting constants
of order 1 (including $\theta$, which is {\em not} assumed to scale).
Here $\varphi$ is of order one. The resulting eigenvalue distribution 
is governed by $V(\varphi)$ \cite{BIPZ}, and it will
remain near the Gaussian fixed
point resp. Wigner's law provided the bare couplings satisfy
$g_n \leq \cN/\a_0^{n} \sim \L^\d$, where
\be
\d = d+n-\frac{nd}2
\ee 
is just engineering dimension of $g_n$. 
As usual, this means that relevant or marginal couplings 
with $\d \geq 0$ are ``safe'' and expected to be renormalizable, while
irrelevant couplings with $\d<0$ must be fine-tuned and
are expected to be non-renormalizable.

In general, it will be  safe to
use \eq{tilde-S} as long as the shape of the resulting eigenvalue
distribution $\obar\varphi(s)$
is close to the Wigner law. For a qualitatively different
shape, one should expect corrections to \eq{tilde-S}. 

In particular, it is quite clear that turning on some small coupling $g_4$ 
must be compensated by a suitable ``mass renormalization'' in order to
preserve the shape of $\obar\varphi(s)$. This allows to determine
the mass renormalization, which will be
explored in Section \ref{sec:weak-coupling}.
But before discussing these issues, let 
us look at the above results from a different perspective:

\subsection{Angle-eigenvalue coordinates in field space}
\label{sec:angle-eigenvalues}

One of the merits of the NC field theory is that it 
naturally suggests new coordinates in field space, which are
very different from the usual ``local'' fields $\phi(x)$ in the
commutative case. This is particularly obvious using a
regularization in terms of finite-dimensional matrices 
$\phi \in Mat(\cN, \C)$. Then there is a natural action of the
$SU(\cN)$ group $\phi \to U^{-1} \phi U$, which 
can be seen as NC version of the symplectomorphisms. 
Even though this is not a symmetry due to the kinetic term,
 it suggests to parametrize $\phi$ in terms of eigenvalues and
``angles'', which are very non-local
coordinates. This change of variables 
 leads very naturally to the picture we found
above. In particular, the
non-trivial measure factor in the path integral due to the 
Jacobian 
makes the existence of a non-trivial eigenvalue distribution very plausible.

We start from the simple fact  that
any hermitian matrix $\phi$ can be diagonalized,
\be
\phi = U^{-1} (\phi_i) U
\ee
where $U \in SU(\cN)$ and $(\phi_i)$ is a diagonal matrix,
with eigenvalues $\phi_i$. We can moreover
assume that the eigenvalues of $(\phi_i)$ are ordered; then 
the matrix $U$ is unique up to phase factors
$K \cong U(1)^{\cN-1}$, provided the eigenvalues $\phi_i$ are non-degenerate. 
This leads to the following definition of the orbits
\be
\cO(\phi) := \{U^{-1} (\phi_i) U; \;\; U \in SU(\cN) \} \cong SU(\cN) / K
\ee
where $K$ is the stabilizer group of $(\phi_i)$. These are compact 
homogeneous spaces.
Then the partition function can be written as
\bea
Z &=& \int \cD \phi\; \exp(-S(\phi))) 
 = \int d\phi_i \D^2(\phi_i) \int dU \exp(-S(U^{-1}(\phi_i) U))  \nn\\
 &=&  \int d\phi_i dU \exp( \sum_{i\neq j}
  \log|\phi_i-\phi_j| - S(U^{-1}(\phi_i) U))  \nn\\
 &=&  \int d\phi_i \tilde F(\phi_i)  
   \exp( -(2\pi\theta)^{d/2}\;\sum_i V(\phi_i) 
  + \sum_{i\neq j} \log|\phi_i-\phi_j|),
\label{Z-noJ1}
\eea
where $dU$ is the integral over $\cN\times \cN$ unitary matrices;
similar manipulations were also done in \cite{martin}.
We introduced here the function
\beq
\tilde F(\phi) := \int dU \exp(-S_{kin}(U^{-1}(\phi) U))  
= : e^{-\tilde\cF(\phi)},
\label{F-def}
\eeq
which by definition depends only on the eigenvalues of $\phi$.
The last form  is justified because $\tilde F(\phi)$ is positive. 
It satisfies
\be
\tilde\cF(\phi+c) = \tilde\cF(\phi), \quad \tilde\cF(-\phi) = \tilde\cF(\phi).
\label{F-shift}
\ee
Moreover, $\tilde\cF(\phi_i)$ is {\em analytic} in the 
$\phi_i$ because the space is compact,
and invariant under exchange of the eigenvalues.
We can therefore expect that it approaches some nice classical
functional  $\tilde\cF[\phi(s)]$ in the large $\cN$ limit, where 
$\phi(s)$ is the function in {\em one} variable which is related to
$\phi_i$ as in \eq{varphi-function}.

We can now read off the induced action for the eigenvalues,
\be
\tilde S(\phi_i) = \tilde\cF(\phi_i) - \sum_{i\neq j} \log|\phi_i-\phi_j|
+(2\pi\theta)^{d/2}\;\sum_i V(\phi_i) 
\label{S-eff-orbit}
\ee
The $\log$ - term in \eq{S-eff-orbit} could also be absorbed by defining
\be
\cF(\phi_i) = \tilde\cF(\phi_i)- \sum_{i\neq j} \log|\phi_i-\phi_j|.
\ee
In particular, note that 
the log-term in \eq{S-eff-orbit} strongly 
suppresses degenerate eigenvalues, and
this cannot be compensated by any analytic $\tilde\cF(\phi_i)$. 
Therefore the saddle-points of 
$\tilde S(\phi_i)$ corresponds to some non-degenerate eigenvalue
distribution. Furthermore the kinetic term (encoded in $\tilde\cF(\phi_i)$)
strongly suppresses jumps in this eigenvalue distribution\footnote{At
  strong coupling however, we will find a phase transition 
to a distributions with one gap, corresponding to some ``striped'' phase},
therefore we expect it to approach some smooth
function $\obar\phi(s)$ in the limit $\cN \to \infty$.
Furthermore, we expect this eigenvalue
distribution to have compact support after a suitable rescaling:
\be
\obar\phi(s) = \a \obar\varphi(s)
\label{alpha-general}
\ee
where $\a$ denotes the maximal eigenvalue.
This typically happens for matrix models, and appears to
be true also in this context as shown below. Indeed 
since $\tilde\cF(\phi)$ only depends on the eigenvalues,
one can trivially interpret it as a function of any hermitean matrix
$\phi = U^{-1} (\phi_i) U$, and rewrite the partition
function as an {\em ordinary} matrix model with formal $U(\cN)$ symmetry,
\be
Z =  \int d\phi_i \exp(-\tilde S(\phi_i)) 
 = \int \cD \phi
   \exp(-\tilde \cF(\phi) -(2\pi\theta)^{d/2}\;Tr V(\phi)).
\label{Z-F}
\ee
The last step is of course completely formal, and we are only allowed
to determine observables depending on the eigenvalues with this action. 
These are determined by the ``effective action'' 
$\tilde S(\phi_i)$, since  the degrees of freedom
related to $U$ are integrated out.  
This strongly suggests that much of the information
about the quantum field
theory, in particular the phase transitions and the
thermodynamic properties, are determined
by $\tilde S(\phi_i)$  and  the resulting eigenvalue distribution 
in the large $\cN$ limit. Note that this is essentially a
one-dimensional problem, governed by the (unknown) functional 
$\cF[\phi(s)]$ and $V$.
The advantage of this
formulation is that it is very well suited to include interactions, 
and naturally extends to the non-perturbative domain. 

Now assume we know
$\cF(\phi_i)$; one can then look for the saddle-points of $\tilde S(\phi_i)$,
determined by
\be
\frac {\d\tilde\cF}{\d \phi_i} + (2\pi\theta)^{d/2}\; V'(\phi_i) 
= \sum_{j\neq i} \frac 1{\phi_j -\phi_i}
\label{EV-saddle-point}
\ee
and ask if they are localized enough to dominate the observables.
If the existence of a sharp eigenvalue
distribution  is established, the full
path integral in \eq{Z-noJ1} would be dominated by the integral over the
corresponding $SU(\cN)$ orbit $\cO(\phi_i)$. This in turn should allow
to recover also other properties such as correlation functions, by
integrating over the corresponding $\cO(\phi_i)$ which is compact. 
This will be discussed further in Section \ref{subsec:physics}.

We can now relate this to the results of Section \ref{subsec:free distri}: 
All observables of the eigenvalues as considered
there are determined by the ``effective action'' $\tilde S(\phi_i)$ above.
The result of Sections \ref{subsec:free distri} 
and \ref{subsec:interactions}, in particular the
factorization property \eq{cluster}, 
says that there is indeed a well-defined eigenvalue distribution in the
non-commutative domain\footnote{
This is not true for conventional field theory
with $\theta=0$, even though the formulation of this section is still
possible for e.g. the commutative limit of fuzzy spaces.}, 
i.e. as long as $\L^2 \theta \gg 1$.
Comparing with the  effective action for the free case
\eq{eff-act-eig},
we find
\be
\tilde \cF(\phi) = f_0(m) + \frac{2\cN}{\a_0^2} Tr \phi^2
\label{F-explicit}
\ee
which certainly reproduces all admissible observables for
 the potential $V(\phi) = \int \frac 12 m^2 \phi^2$. 

This formula may appear strange, since
$\tilde \cF(\phi)$ should of course be independent
of $V$. The reason is that this relation \eq{F-explicit} 
has been established only
``on-shell'', for eigenvalue distributions close to the
Wigner law for the free case. 
It is not clear how well this works for large deviations from that
case.
Later we will use this form also for eigenvalue
distributions which are quite different from the free one, where
one should expect corrections to \eq{F-explicit}. 
The appropriate way to use \eq{F-explicit} for some given eigenvalue 
distribution is therefore
to determine $m$ such that the corresponding
free distribution matches best the one under
consideration.
It would be extremely interesting to know more about the  functional
$\tilde\cF(\phi)$.

The dominance of
a given orbit $\cO(\phi_i)$ in \eq{Z-noJ1} is clearly related to
 UV/IR mixing: naively, the action has a minimum
at $\phi =const\; \one$; however if the volume-factors from the path 
integral are taken into account (which happens at one loop), 
this zero-momentum state  
 is actually highly suppressed, and the dominating field configurations 
have nontrivial position-dependence (i.e. momentum), 
due to the nontrivial eigenvalue
distribution. Note that this argument is completely 
non-perturbative. 
Furthermore, depending on the form of the potential $V(\phi)$ 
the dominating eigenvalue distribution may be connected or consist of
disjoint pieces. These would clearly correspond to different
phases. This picture will be made more quantitative below, and 
we will be able to identify the ``striped'' phase of \cite{gubser}.

\subsubsection{Interpretation of the orbits  $\cO(\phi_i)$.}
\label{subsubsec:interpret}

Consider the reduced model \eq{F-def} for a given orbit $\cO(\phi_i)$
with fixed eigenvalue distribution, whose (free) energy is given by 
$\tilde \cF(\phi)$.
Intuitively, we can interpret this model as follows: 
consider a classical fluid $\phi(x)$ on a compact space 
(due to the regularization e.g. on $\C P^n$ or some torus) 
with prescribed ``density'' 
\be
\rho(p) = \frac 1V \int d^d x\; \d(\phi(x) - p),
\label{class-constraint}
\ee 
corresponding to the 
eigenvalue distribution. Then $\rho(p)$ is essentially the 
density of eigenvalues \eq{density-phi}, at least in the semi-classical
limit. Note in particular that the action of the classical volume-preserving 
diffeomorphisms on $\phi(x)$ can be approximated by $SU(\cN)$, since
any configuration with given $\rho$ can be obtained using $SU(\cN)$.
Entropy will favor mixing, but the ``kinetic energy''
suppresses mixing. 
In a small region of space, the global constraint
\eq{class-constraint}
is quite irrelevant, and the fluid will behave like a fluid with the
same action but without the constraint. However, if we 
fix the field on a large part of the volume, this must be
compensated in the remaining space. 
This is clearly an IR effect, suppressing very large wavelengths. 
Therefore we expect that 
the theory  behaves like an
``ordinary'' field theory in small enough regions of space
and may hence describe ordinary local
physics, however it is certainly different globally. This is quite
interesting and encouraging 
for possible applications in elementary particle physics.

In order to go beyond the 
computations in the following sections, 
one should therefore study the reduced models \eq{F-def} in more
detail, and see to what extent they approximate
a scalar field theory. 
Note that quite generally if the scale $\a$ is increased, 
the configurations with 
short wavelength will be more strongly suppressed, leading a long correlation 
length; on the other hand for small $\a$, 
the correlation length will be short. 
Therefore there should indeed exist some suitable scaling $\a(\cN)$ which gives
a macroscopic correlation length in the limit $\cN \to \infty$. 
The relation with the intuitive picture 
presented in Section \ref{sec:angle-eigenvalues} and in particular 
\eq{class-constraint} can be seen best using coherent states 
or projectors (see
e.g. \cite{vaidya,cpn-bala,szabo-lizzi}), in 
particular for the regularizations with $\C P^n_N$.
It would be very interesting to combine these methods with the approach in
the present paper.

The ``ground state'' of \eq{F-def}
on a given orbit $\cO(\phi_i)$ is some very smooth
function with the given density, which solves the e.o.m.
\be
[\D\phi,\phi] =0.
\label{orbit-saddle}
\ee
This has many interesting solutions: One class is
given by solutions of the free wave equation $\D\phi = c\phi$. 
In particular, the (non-commutative) spherical harmonics with suitable
eigenvalue density are solutions also on the above orbits $\cO(\phi_i)$.
However there are other solutions, for example any solution of
$\D\phi = f(\phi)$ for arbitrary $f$ solves \eq{orbit-saddle}; in
particular any diagonal matrix does (in the usual basis). 
A careful study of these issues
should lead to a better understanding of \eq{F-def}, and hence to
improvements of the simple results presented below.

\subsection{Relating the matrix model to physical observables}
\label{subsec:physics}

Before analyzing further the matrix models \eq{tilde-S}, we
should try to relate them to the physically interesting quantities such as
mass and coupling constants. Recall that the 
definition of mass on a non-commutative space is not obvious,
since the Lorentz-invariance is generally broken. 
However on Euclidean $\R^{2n}_\theta$
with $Sp(2n)$- invariant $\theta_{ij}$ one can define 
a {\em correlation length} in terms of correlation functions for 2 
suitably localized Gaussian
wave-packets\footnote{These
``test-functions'' can be moved transitively on the NC spaces using
the translational symmetry which is unbroken. On 
$\R^{2n}_\theta$ with $U(n)$- invariant $\theta_{ij}$ 
resp. their fuzzy versions $\C P^{n}_N$,
the residual unbroken $U(n)$ rotational symmetry is maximal and
large enough to ensure 
that the correlation length is independent of the direction.} 
$\phi(x)$ of size $1/\sqrt{\theta}$, say, 
or  e.g. coherent states for the fuzzy spaces $\C P^n_N$:
\be
\langle \phi(x_1)\;\phi(x_2)\rangle = \int d \phi_i \D^2(\phi_i)
\int_{\cO(\phi_i)} dU e^{-S(U^{-1} (\phi_i) U)}
\langle\phi(x_1),U^{-1}(\phi_i) U\rangle 
\langle\phi(x_2),U^{-1}(\phi_i) U\rangle.
\label{correl-def}
\ee
Here $\langle\phi(x),U^{-1}(\phi_i) U\rangle \propto Tr
(\phi(x)U^{-1}(\phi_i) U)$ denotes the inner product on the NC
space. Note again that only the kinetic part of the action is non-trivial
here, and we will assume in the following
that the full path integral is dominated by some orbit
$\cO(\phi_i)$.
The mass can then be identified as the inverse correlation length, 
provided   it is $\ll m_\theta$. 
Similarly, one could define the
coupling constants in terms of correlation functions of e.g. 4 such
wave-packets.
In general, it is plausible that the interesting low-energy observables
depend only on the eigenvalue
distribution, and can be determined in principle 
by an integral over the orbit $\cO(\phi_i)$.

The question of renormalizability is then roughly
whether one can scale the (finitely many) 
couplings $g_n$ with the cutoff $\L \to
\infty$ in such a way that the correlation length and
all the other  observables at low energy 
(or at the scale of non-commutativity set by $\theta$) 
approach a well-defined limit.
To answer this question of course requires control over all these
correlation functions. In this paper, we try to proceed as much
as possible without resorting to perturbation theory.
In view of the above results on the eigenvalue
distribution, it seems very plausible that the scaling of the
bare couplings $m$ and $g_n$ must be determined such that {\em the shape
of the dominating eigenvalue distribution, i.e. the normalized function 
$\obar\varphi(s)$ \eq{alpha-general} is fixed as $\L \to \infty$}. 
These scalings should be easily
accessible with matrix model techniques if we know the matrix model,
in particular $\tilde\cF(\phi)$. Moreover, it seems plausible that
this should even be sufficient to guarantee ``renormalizability'' 
in the above sense, provided the field theory can indeed be
reduced to the orbit $\cO(\phi_i)$ as discussed in Section 
\ref{sec:angle-eigenvalues}.

To make explicit computations, we have to use \eq{tilde-S} 
resp. \eq{F-explicit} for the time
being, i.e. we have to require that
$\obar\varphi(s)$ 
respectively its related eigenvalue density \eq{density-phi}
is close enough to the Wigner law \eq{wigner-law}.
Then the ``physical'' mass $m_R$ resp. correlation
length can be identified
without computing such correlation functions:  it should be given by
the bare mass of the corresponding free theory 
{\em with the same
maximal eigenvalue $\a_0(m_R)$}, for the same cutoff $\L$.
That is, $m_R$ is determined by
\be
\a = \a_0(m_R)
\label{alpha-mass}
\ee
where $\a$ \eq{alpha-general} will depend on the couplings 
of the interacting matrix
model. This will
be elaborated in Section \ref{sec:phi4}. We will apply this
prescription \eq{alpha-mass} even if the function
$\obar\varphi(s)$ 
is not close to the free one in this paper, 
hoping that it is mainly the ``size'' of the eigenvalue distribution
which determines the correlation length. This is plausible in
view of the classical picture discussed in Section 
\ref{subsubsec:interpret}.

We want to point out again the following consequence of this picture:
 \eq{correl-def} is strongly suppressed  for zero momentum
$\phi_0 \propto \one$,  since
$\langle\phi_0,U^{-1}(\phi_i)U \rangle =0$ for the
dominating eigenvalue-distribution. 
On the other hand for non-zero momentum, the
eigenvalues are non-degenerate, and the above inner product is
non-zero.  This suppression of zero momentum 
is clearly related to  UV/IR mixing, and suggests that indeed only
localized wave-packets should be used.

\section{Example: the $\phi^4$ model}
\label{sec:phi4}

Consider now the model 
\bea
S &=& \int d^d x \(\frac 12 \partial_i \phi \partial_i \phi 
+ \frac 12 m^2 \phi^2 + \frac{g}{4} \phi^4\) \nn\\
&=& \int d^d x \(\frac 12 \partial_i \phi \partial_i \phi 
+ \frac 12 m_R^2 \phi^2 + \( \frac 12 \D m^2 \phi^2 + \frac{g}{4}
\phi^4\)\) 
\label{phi4-renorm}
\eea
where we have introduced a ``renormalized'' mass $m_R^2 = m^2 - \D m^2$.
Its eigenvalue sector according to the above results 
is described by 
\be
\tilde S = f_0(m_R) +\frac{2\cN}{\a_0(m_R)^2}\; Tr \phi^2
+(2\pi\theta)^{d/2} Tr\( \frac 12 \D m^2 \phi^2 + \frac{g}{4}\phi^4\).
\label{phi4-matrix}
\ee
While $m_R$ is arbitrary in principle, we will adjust it in order  to
minimize the expected errors due to our only partial knowledge of
$\cF(\phi_i)$. 
Note that 
in the regularization using fuzzy $\C P^n$, we will find \eq{cutoff-CPn}
\be
\L  = \sqrt{\frac{2 N}{\theta}}
\label{L-CPN}
\ee
where $N$ is related to $\cN$ via \eq{n-cN-CPn},
and a similar result 
$\L  = \sqrt{\frac{\pi N}{\theta}}$
using NC tori \eq{NC-lattice-cutoff} where
$\cN = N^{d/2}$. Therefore
\be
\frac{\cN}{\a_0^2} = O(N \theta^{(d-2)/2}) = O(\L^2 \theta^{d/2})
\ee
up to $\log$- corrections.

To solve this model, we first rescale $\phi$ as 
\be
\phi = \a_g \varphi
\ee
such that the saddle-point solution for
$\varphi$ will have an eigenvalue distribution with 
range\footnote{in the disordered phase discussed at present, see
  Section \ref{subsec:phase1}} $[-1,1]$.
Rewriting the matrix model \eq{phi4-matrix} as
\be
\tilde S =  f_0(m_R) + \cN Tr \(\frac{m'^2}{2}\varphi^2
 + \frac{g'}{4} Tr \varphi^4\).
\label{phi4matrix-model}
\ee
we have
\bea
\a_g^2\(\frac{2\cN}{\a_0(m_R)^2} + (2\pi\theta)^{d/2} \frac 12 \D m^2 \)
  &=& \frac{\cN} 2 m'^2,   \label{eq-ren-1}\\
(2\pi\theta)^{d/2} g \a_g^4&=& \cN g',    \label{eq-ren-2}\\
m^2 &=& m_R^2 + \D m^2 .         \label{eq-ren-3}
\eea
The model \eq{phi4matrix-model} is well-known
 and can be solved with 
standard methods from random matrix theory, see
e.g. \cite{BIPZ,eynard}. It is again governed by a 
single saddle-point in the eigenvalue sector, with eigenvalue distribution 
$\obar\varphi_g(s)$ resp. $\rho_g(p)$. Note that now $m'^2 <0$ is
allowed. Assuming that $g'>0$, there are 2 cases corresponding to
distinct phases of the model which we are discussed below. 

It is interesting to note that even small $g'<0$ is admissible
for this matrix model as long as $m'^2>0$. This 
indicates analyticity in $g'$, so that perturbation
theory does make sense in the weakly coupled phase.

\subsection{Phase 1 (``single-cut''): $\;m'^2>0$, or $m'^2<0$ 
with $\frac{m'^4}{4 g'}<1\;\;\;$}
\label{subsec:phase1}

In this case the eigenvalue density is given by \cite{BIPZ}
\be
\rho_g(\varphi) = \frac 1{2\pi} (g' \varphi^2 + \frac{g'}2 +m'^2)
 \sqrt{1^2-\varphi^2}.
\label{distri-int}
\ee
We have imposed that the 
range of $\varphi$ be $[-1,1]$ as explained above, which using 
standard formulas \cite{BIPZ} implies
\be
1 = \frac{2}{3g'}(-m'^2+\sqrt{m'^4+ 12 g'})
\label{support-distri-int}
\ee
i.e.
\be
g' = \frac 43(4-m'^2)
\label{gprime-mprime}
\ee
Note that the matrix model \eq{phi4matrix-model}
has 2 independent parameters, which can be chosen either as $m'^2$ and
$g'$, or e.g. $\a_g$ and $g'$. We have chosen to work with $\a_g$, 
therefore $m'^2$ and $g'$ are not independent.
In particular, note that $\frac{m'^4}{4 g'}<1$ for $g' < 16$ 
due to \eq{gprime-mprime}, therefore this phase will be the 
``weakly-coupled'' phase in Section \ref{sec:weak-coupling}.

Inserting $m'^2 = 4-\frac 34\; g'$ in \eq{eq-ren-1} 
and using \eq{eq-ren-2} gives
\bea
&&\frac{2\a_g^2}{\a_0^2} + (2\pi\theta)^{d/2} \a_g^2\frac 1{2\cN} \D m^2
 = 2-\frac 38\; (2\pi\theta)^{d/2}\frac{\a_g^4}{\cN}\; g.
\label{matching-2}
\eea
So far, $m_R$ was arbitrary, and it shouldn't matter 
for fixed ``bare parameters'' $m$ and $g$.
However, the relation \eq{tilde-S} is expected to be good only if the
eigenvalue distribution is close to the free one corresponding to
$m_R$. We should therefore  choose $m_R$ accordingly, and
a good choice is
\be
\a_0(m_R) = \a_g
\ee
as in \eq{alpha-mass}. This guarantees that the eigenvalue
distribution of the interacting model \eq{phi4matrix-model} has 
the same range as
the free one obtained from $m_R$, hence it is ``close''.
Then \eq{matching-2} simplifies further, and 
using $\D m^2 = m^2 - m_R^2$ we have 
\bea
(2\pi\theta)^{d/2} g\; \a_0^4(m_R)&=& \cN g' \label{g-match}\\
m^2 +\frac 34 g\;  \a_0^2(m_R)  &=& m_R^2.  \label{eq-ren-3new}
\eea
These are 2 equations in 4 unknowns (the function $\a_0(m_R)$ being
known \eq{alpha}), and we can basically choose any 2
of them to parametrize the model.

\subsection{Phase 2 (``2 cuts''): $\;m'^2<0$ with $\frac{m'^4}{4 g'} >1\;\;$}

At $4g' = m'^4$ with $m'^2<0$
 the eigenvalue density breaks up into 2  disjoint peaks,
which are concentrated around the 2 minima.
 For $4g' <  m'^4$ these peaks have finite distance,
and moreover spontaneous symmetry breaking of $Tr \phi$ occurs 
with non-zero ``occupation'' in both peaks
\cite{cicuta-magnet}.  
This clearly describes a different phase of the model.
The semi-classical interpretation in the picture of Section 
\ref{subsubsec:interpret} would be a 2-fluid model, with
a large surface energy at the contact surface due to the kinetic term. 
It is very plausible that this will develop some kind of (generalized)
striped pattern upon mixing, and we conjecture this to be the
``striped'' phase of \cite{gubser}, or equivalently the ``matrix
phase'' of \cite{martin}. Note that the assumption in 
\cite{martin} of 2 delta-like peaks is qualitatively very close, but
strictly ruled out by the log-terms in \eq{S-eff-orbit}.

It is even possible to consider negative $g'$ in this
model \cite{BIPZ}, reflecting the analyticity in $g'$. If $g'$ becomes
too negative there is another phase transition, 
which will not be considered any further here.

\section{Weak coupling and renormalization}
\label{sec:weak-coupling}

We first consider the weak coupling regime, and try to get some insights
into the renormalization of the $\phi^4$ model. 
In particular, 
one would like to understand better the relation of the above approach
with the usual concepts of running coupling constants etc. 
The natural scaling parameter here is the size of the matrices
$\cN \sim N^{d/2}$, which is  related to the UV-cutoff $\L$
through \eq{L-CPN} for the regularization using fuzzy $\C P^n$. 
We can therefore ask how the bare parameters 
$m^2, g$ etc. must  be scaled with $\L$ such that 
the ``low-energy physics'' remains invariant, and we could define the 
corresponding beta-functions.

In the weak coupling regime, the 
shape of the eigenvalue distribution  $\obar\varphi$
\eq{alpha-general} resp. $\rho_g$ \eq{distri-int}
should certainly be kept
close to the free one, given by the Wigner law. 
As discussed in Section \ref{subsec:physics},
 the correlation length resp. ``renormalized mass'' $m_R$
can then be obtained simply
by matching the size of the eigenvalue
spectrum $\a_g$ with the one for the free case with the same cutoff, 
\be
\a_g = \a_0(m_R)
\ee
see \eq{alpha-mass}.
Then $m_R$ measures the correlation length of the free
model which best fits the eigenvalue distribution.
This should be very reasonable, 
since we are looking at a weak coupling expansion
where $g'$ can be taken arbitrarily small.

The corresponding running of the original parameters $m$ and $g$
depends on the dimension, and will be worked out below.
For the mass, this is very easy in our approach. 
Unfortunately, it is not obvious how to relate the coupling $g$ 
resp. $g'$ to e.g.  4-point
functions or scattering processes; this could be supplemented
perturbatively by a 
conventional RG computation, cp. \cite{wulki,pietroni}. 
Also, note that there is no  wavefunction-renormalization in this
approach, as there are only 2 coupling constants in the matrix model 
\eq{phi4matrix-model}. 
On the other hand the computations below are much simpler than any
standard field-theoretic computations, and provide
non-trivial information 
on the model without having to resort to perturbation theory.

\subsection{2 dimensions}

To find the scaling of $m^2$ and $g$ we can use \eq{g-match}
and \eq{eq-ren-3new}, where $\a_g = \a_0(m_R)$ is already built in. 
The running of the mass can be read off from \eq{eq-ren-3new},
\be
m^2  =  m_R^2 - \frac 34 g  \a_0^2(m_R) 
  = m_R^2 - \frac 3{4 \pi} g \ln(1+\frac{\L^2}{m_R^2})  
\label{mass-ren-2d}
\ee
Recall that $m_R$ is the ``physical'', renormalized mass which is
supposed to be finite while $\L \to \infty$. We see explicitly the 
expected logarithmic divergence of the bare mass. Note that this is an 
all-order result in $g$, despite appearance. 

It is interesting to compare this with a conventional one-loop
calculation. 
Noting that 
our interaction term is $\int \frac g4 \phi^4$ 
(which is more natural in the matrix
model context) rather than $\int \frac g{4!} \phi^4$, one would find
\be
\d m^2_{P + NP} = 12\; \frac{g}{4} \int \frac{d^2 k}{(2\pi)^2} \frac
1{k^2+m_R^2} =  \frac{3}{4\pi} g \ln(1+\frac{\L^2}{m_R^2})
\ee
from the $8$ planar plus $4$ non-planar contractions.
It is interesting to see that this one-loop computation
agrees precisely with our procedure of
matching the eigenvalue distribution with the free one,
even though we used the non-perturbative matrix model results 
\eq{distri-int} ff. One might have
expected that only the planar diagrams contribute to the 
mass renormalization; however since the
``mass'' is obtained as the $p \to 0$ limit of the 1PI 2-point
functions (for fixed $\L$), 
the planar and non-planar diagrams coincide and both contribute. 

Note that if we use another regularization such as NC tori, 
we should use the modified propagator  \eq{lattice-propagator}  
rather than the sharp cutoff in \eq{phi2-expect-2d}. 
This would lead to a somewhat modified formula for
$\a_0(m_R)$, however the corresponding mass renormalization 
\eq{mass-ren-2d} would still coincide with the one-loop result 
since the same $\a_0(m_R)$ enters in both calculations.

Next consider the bare coupling $g$. 
Since the matrix model provides no simple relation between $g$ and 
the 4-point function, the latter would have to be obtained 
e.g. by a perturbative calculation as usual. 
However since $g$ is not expected to run in 2
dimensions, we simply interpret $g$ as ``physical'' coupling constant,
fixing the scale by $m_\theta$. 
The relation with the matrix model coupling $g'$ is given by
\eq{g-match}, which for large $\L$ and using \eq{L-CPN} is
\be
 g\; = \frac{\L^2 \pi}{16(\ln(\frac{\L}{m_R}))^2 }\; g'.
\label{g-match-2d}
\ee
In particular if we keep $g$ finite,
we see that $g' \to 0$ very rapidly as $\L \to \infty$. 
This means that the eigenvalue density $\rho_g$ \eq{distri-int} 
is very close to
the free one given by Wigner's law, and we expect that the above
relations are reliable in this
weak-coupling phase of $\phi^4$ in 2 dimensions\footnote{ 
Recall however that the
non-planar diagrams are suppressed only logarithmically 
\eq{planar-approx}  in 2D, which is much weaker than in 4D}.

\subsection{4 dimensions}

Using the same procedure, \eq{eq-ren-3new} gives now
the expected quadratic divergence of the bare mass
\be
m^2  =  m_R^2 - \frac{3}{16\pi^{2}}\;
   \L^2 \(1 -\; \frac{m_R^2}{\L^2}\;\ln(\frac{\L^2}{m_R^2})\)\; g.
\label{mass-ren-4d}
\ee
This is again an all-order result, and agrees again perfectly with
a conventional one-loop computation (see e.g. (3.11) of \cite{seiberg}
after replacing $g \to g/6$). 
However the bare coupling $g$ is now
also expected to run, being logarithmically divergent at one loop.
Recall also that we cannot see any
wavefunction-renormalization as there are only 2 parameters in the
matrix model. The relation between $g$ and the
matrix-parameter $g'$ is now given by
\be
g' = \frac{2}{\pi^2} 
\(1 -2\; \frac{m_R^2}{\L^2}\ln(\frac{\L}{m_R})\)^2 \; g
 \approx \frac{2}{\pi^2} \; g
\ee
for $\L \gg m_R$,
where \eq{L-CPN} and $\cN = N^2/2$ \eq{n-cN-CPn}
has been used. This means that (keeping in mind some possibly
logarithmic renormalization of $g$) our procedure to derive \eq{mass-ren-4d}
is justified in a perturbative sense, since we can make
$g'$ arbitrarily small so that again $\rho_g$ 
is very close to Wigner's law.

It is quite remarkable that according to the above results, 
the mass is renormalized only at first order in the
bare coupling $g$, even though the derivation is non-perturbative. 
While this is expected for $d=2$, this seems very surprising
for $d=4$; it would indicate that NC $\phi^4$ in 4 dimensions is much
better behaved than in the commutative case. 
This can be traced back to our procedure
of determining the mass $m_R$ through $\a = \a_0(m_R)$. 
While an exact calculation of
the correlation length on a given orbit $\cO(\phi)$
would certainly modify this result, the basic
picture seems clear and simple. Unfortunately this
analysis does not give us the relation between $g$ and the
``physical'' coupling as obtained by the 4-point
function, which probably requires further renormalization.

It is tempting here to conjecture that fixing $g'$ 
and $m'$ or equivalently the eigenvalue distribution $\obar\varphi$
in a suitable way
suffices to define a non-trivial NC field theory in 4 dimensions. 
We will find further evidence for this in Section \ref{subsec:4dcrit}.

We conclude that while this approach certainly needs further work
and thus far provides only a partial window into NCFT, it 
suggests a very simple 
and compelling approach to renormalization in the NC case.
This may help to overcome the difficulties at
higher order found in \cite{seiberg}.

\subsection{Higher dimensions}

It is also illuminating to try to generalize the above considerations
to dimensions higher than 4, where the commutative $\phi^4$ model is no longer
renormalizable. The relations \eq{eq-ren-3} and \eq{g-match} are still
valid, and give
\be
m^2 
 =  m_R^2 -  g \; O(\L^{d-2})
\label{mass-ren-d}
\ee
and
\be
g' = O(\L^{d-4})\; g.
\label{g-ren-d}
\ee
Clearly our matrix model \eq{phi4matrix-model} makes sense 
only for $g' =O(1)$, which would require $g = O(\L^{4-d})$. This is just the
canonical dimension of $g$. If we again assume that $g'$ 
and hence the shape of the eigenvalue distribution should be
independent of $\L$, we see that $g$ must be fine-tuned to at least
$O(\L^{4-d})$.
This is in accord with the expected non-renormalizability.

\section{The phase transition.}
\label{sec:phase-transition}

Now consider the phase transition of the
matrix model \eq{phi4matrix-model}, which is known 
\cite{shimamune,eynard} to have a 
phase transition between the 1-cut and the
2-cut phase  at 
\be
m'^4 = 4 g'.
\label{phasetransition-1}
\ee
This is expected to be the phase transition between the 
disordered and striped phase of Gubser and Sondhi.
Combining with \eq{gprime-mprime} this gives
\be
g' = \frac 43(4-m'^2) = \frac 14 m'^4 
\ee
which has 2 solutions for $m'^2$; only the negative one 
\be
m'^2  =-8 
\ee
is relevant here and marks the phase transition.
The corresponding coupling is\footnote{Recall that $g' < 16$ 
is the weakly coupled single-cut phase as pointed 
out in section \ref{subsec:phase1}.}
\be
g' = 16.
\ee
Note that there is still a free parameter in the model, which we can
take to be $m_R$ resp. $\a$. Plugging this in 
\eq{g-match} gives
\bea
(2\pi\theta)^{d/2} g \a_0^4(m_R) &=& 16 \cN, \label{galphaN}\\
m^2 +\frac 34\;g\;  \a_0^2(m_R)  &=& m_R^2,  \label{eq-ren-4}
\eea
and solving the first equation for $\a_0$ we obtain
\be
m^2 + 3\;\sqrt{\frac{\cN g}{(2\pi\theta)^{d/2}}}  = m_R^2(g),
\label{eq-ren-5}
\ee
where $m_R^2$ is a function of $g$ via \eq{galphaN}.
This defines the critical line.
The corresponding phase transition is third order 
\cite{shimamune}
in the variables $g'$ and $m'$. 
This is in contrast with \cite{gubser,wu,zappala}, 
who argued for a first-order phase transition.

Note that $m'^2<0$ means that we are quite far
from the perturbative domain, and the eigenvalue
distribution is significantly changed from 
Wigner's law \eq{wigner-law}. In this regime the
replacement \eq{eff-act-eig} for the
kinetic term has not been tested, and we cannot expect the results
to be exact. 
Furthermore, we expect the ``bare''
parameters $m, g$ to have some non-trivial scaling in $\cN \sim N^{d/2}$
on the critical line corresponding to the RG flow, which is not obvious.
In any case, it is quite reasonable that this
description is at least qualitatively correct. 
To proceed, we have to discuss the dimensions separately, starting
with the most interesting case of 4 dimensions.

\subsection{4 dimensions}
\label{subsec:4dcrit}

\begin{figure}[htpb]
\begin{center}
\epsfxsize=3in
   \epsfbox{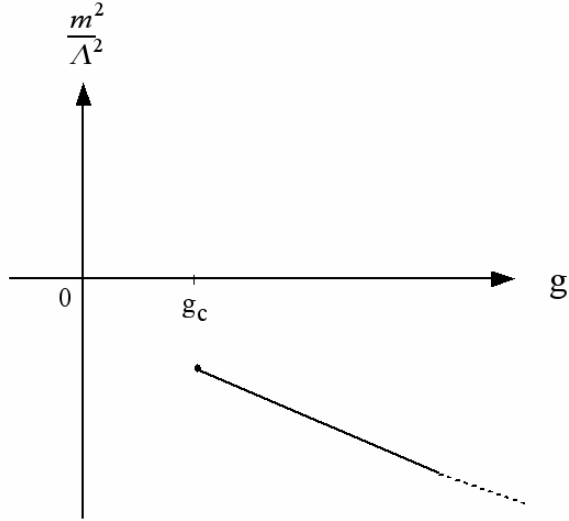}
\end{center}
 \caption{The critical line in 4D with critical point}
\label{fig:critical4d}
\end{figure}

We consider first the regularization using fuzzy $\C P^2$
discussed in Appendix A, which
corresponds to a sharp cutoff. 
In this case we have $\cN = N^2/2$, and 
$\a_0^2 = O(\L^2)$ in $d=4$ \eq{alpha}.
By looking at the equations \eq{galphaN}, \eq{eq-ren-4} and
\be
\L^2 =  \frac{2 N}{\theta}
\label{La-N-relation-CP2}
\ee
which holds for this regularization of $\R^4_\theta$,
it is quite obvious that
there should be solutions which scale as
\be
m^2 \propto \L^2 \propto N, \quad g \;\;\;\mbox{fixed}.
\label{scaling-4d}
\ee
Hence we expect a phase transition at finite coupling $g$ and the standard
quadratic running of the mass.
To find a closed equation for the critical line, we have to 
use \eq{alpha} for $\a_0 = \a_0(m_R)$,
\be
\frac{4\pi^2\a_0^2(m_R)}{\L^2} = 1-\frac{m_R^2}{\L^2} \ln(1+\frac{\L^2}{m_R^2})
\label{m-R-4d}
\ee
This cannot be solved explicitly for $m_R$, but we can use \eq{eq-ren-5} for
$m_R^2$ which together with $\cN = N^2/2$ and \eq{La-N-relation-CP2} gives
\be
\frac{m^2}{\L^2} + \frac{3}{4\pi}\;\sqrt{\frac{g}{2}}  
 = \frac{m_R^2}{\L^2}.
\label{m_R-4d}
\ee
Since $m_R^2\geq 0$, this makes sense only for 
\be
\frac{m^2}{\L^2} > -\frac{3}{4\pi}\;\sqrt{\frac{g}{2}}.
\ee
Plugging this in \eq{m-R-4d} and using \eq{galphaN} 
\be
\a_0^2(m_R) =  \frac{N}{\pi\theta}\;\sqrt{\frac{2}{g}}  
 = \frac{\L^2}{\pi \sqrt{2g}}
\ee
we get
\be
2\pi\sqrt{\frac{2}{g}}
= 1-\(\frac{m^2}{\L^2} + \frac{3}{4\pi}\;\sqrt{\frac{g}{2}} \)
\ln\(1+\(\frac{m^2}{\L^2} + \frac{3}{4\pi}\;\sqrt{\frac{g}{2}}\)^{-1}\).
\label{m-R-4d-2}
\ee
This is indeed consistent with the scaling \eq{scaling-4d}.
Since the rhs is $\in (0,1]$, this has a solution 
only\footnote{the rhs of \eq{m-R-4d-2} is simply 
$c(\L,m_R)$ as defined in \eq{phi2-expect}, 
which will very generally satisfy such bounds.}
for $2\pi\sqrt{\frac{2}{g}} \leq 1$, i.e.
\be
g \geq g_c = 8 \pi^2.
\ee
In terms of the dimensionless parameter
$\tilde m^2 = m^2/ \L^2$, the corresponding critical mass is
$\tilde m^2_c = -\frac 32$. Expanding $g = g_c + \d g, \;\; \tilde m^2
= \tilde m^2_c + \d \tilde m^2$ the critical line \eq{m-R-4d-2} is given by
\be
\d g = -\frac{32 \pi^2}{3}\; \d\tilde m^2 + ...
\ee
for small variations. This is plotted in Figure \ref{fig:critical4d}.

Let us try to assess the validity of this result.
The basic arguments underlying this result are 
expected to be good in 4 dimensions
as long as $m_R^2 \ll \L^2$ (unlike in 2 dimensions, see below
and Appendix B). This 
is satisfied here since 
as $g \to g_c$ from above, \eq{m-R-4d-2} implies that
\be
\frac{m_R^2}{\L^2} = \frac{m^2}{\L^2} +
\frac{3}{4\pi}\;\sqrt{\frac{g}{2}} \quad \to \;\; 0
\label{m-R-limit}
\ee 
using \eq{m_R-4d}, and in particular $m_R =0$ at $g=g_c$. Therefore the 
replacement of the free action 
with the matrix model \eq{eff-act-eig} is essentially justified, 
apart from the 
modified eigenvalue distribution $\obar\varphi(s)$. 
It would be very interesting to estimate the effects of this modified 
$\obar\varphi(s)$ more rigorously, e.g. by 
estimating the integrals over the compact orbits $\cO(\phi_i)$. 
The existence of $g_c \neq 0$ can also be understood simply by 
noting that the critical line is characterized by a specific
eigenvalue distribution which is different from Wigner's law, which
holds for $g=0$. Therefore the critical line\footnote{note that 
$m^2 \propto \L^2$ here, as in the weakly-coupled regime. This is
different in 2 dimensions, where the critical line is in a different
scaling regime from the weakly-coupled case, and no such conclusion can
be drawn.}
 cannot end at $g=0$. 
While the existence of a critical point should be a sound prediction, the 
relation between $m^2$ and $g$ on the critical line
cannot be expected to be precise.
If we would use the propagator for the
lattice regularization \eq{lattice-propagator}, the details of \eq{m-R-4d-2}
would change but the results would be
qualitatively the same. In particular the critical
coupling could be evaluated 
using \eq{alpha-lattice-4d}.

The existence of a critical point terminating the critical line at 
$g_c\neq 0$ is certainly
intriguing. Since the critical line will be stable under the RG flow, 
its endpoint should correspond to a non-trivial RG fixed point for the
non-commutative $\phi^4$ theory, and the correlation length is
expected to diverge\footnote{we find 
indeed $m_R=0$ at $g_c$,
 which should however not necessarily be taken at face value}. 
That limiting model should be non-trivial, since 
 the eigenvalue distribution $\obar\varphi(s)$ is different from the
 free case. It is furthermore plausible following the discussion in 
\cite{gubser} that it admits a 
``continuum'' interpretation as a renormalizable NC field theory.
The physical content of such a fixed point would require to study
e.g. 4-point functions and the running of the coupling, which is beyond
the scope of this work.
Note however that 
if we assume that $g$ becomes larger with increasing cutoff as in the
commutative case, 
this would mean that the low-energy coupling
corresponding to $g_c$ is small, and hence in a physically
interesting regime.

These results and in particular the existence of a phase transition
are roughly consistent with the 
results of \cite{gubser,wu,zappala}, but not precisely; 
for example \cite{gubser,wu,zappala} argue
for a first-order phase transition using self-consistent
Hartree-Fock approximation resp. a one-loop RG analysis,
while we find a higher-order transition 
using a non-perturbative matrix model result. The approaches are
thus quite orthogonal: in the present approach the interaction is 
treated exactly while the kinetic term is approximated, whereas
\cite{gubser,wu,zappala} do the opposite. The fact that one always
finds a phase-transition is quite encouraging. 
It is also remarkable that $\theta$ does not enter our result
\eq{m-R-4d-2}; recall however that we always assume $\L^2 \theta \gg
1$ and \eq{cutoff-CPn}.
In particular, one would expect a 
standard Ising transition to a uniform symmetry breaking state 
at small $\theta \L^2$; however 
our approach is not valid in that regime, 
and we should therefore not expect to
see this transition.

In dimensions higher than 4,  \eq{galphaN} together with 
$\a_0^2 = O(\L^{d-2}) = O(N^{d/2-1})$ 
implies that $g = O(N^{-\frac d2 +2})$ must be fine-tuned in order to
stay in the weakly-coupled phase. This is consistent with the 
expected non-renormalizability, and we will not pursue this any further.

\subsection{2 dimensions}

In 2 dimensions our approach is more delicate as 
the planar diagrams are only logarithmically divergent. To be safe we
should allow only finite $m_R$ while $\L \to \infty$, see Appendix B.
Using \eq{alpha} and \eq{galphaN} 
for $\a_0 = \a_0(m_R)$ and $\cN = N$, we have
\be
m_R^2 = \frac{\L^2}{e^{\pi \a_0^2}-1} 
 = \frac{\L^2}{e^{2\sqrt{\frac{2\L^2}{g}}}-1}.
\label{m-R-2d}
\ee
Plugging this in \eq{eq-ren-5} and using the relation 
$\L^2 =  2 \frac{N}{\theta}$
for the regularization using the fuzzy sphere gives
\be
m^2 + \frac{3}{2}\;\sqrt{\frac{g }{\pi}}\;\L
= \frac{\L^2}{e^{2\sqrt{\frac{2\L^2}{g}}}-1} = m_R^2.
\label{2d-critical}
\ee
This would be consistent with a scaling 
\be
m^2 \sim \L^2, \quad g \sim \L^2
\label{scaling-crit-2d}
\ee
on this critical line\footnote{note that both $m^2$ and $g$ have dimension
of $mass^2$; nevertheless, this is a rather strange scaling in 2
dimensions. This phase transition is very far from the 
weak coupling phase}. 
However,  the basic assumptions of Sections \ref{subsec:free distri} 
and Appendix B
are no longer valid in this case, since then 
$m_R^2 \propto \L^2$. 
Our approach should be most reliable if $m_R$ remains finite,
which implies through \eq{m-R-2d} that 
$g \sim 2(\frac{\L}{\ln \L})^2$ and therefore \eq{2d-critical}
$m^2 \sim -\frac 3{\pi} \frac{\L^2}{\ln \L}$. 
Since we cannot solve \eq{2d-critical} in closed form, let us assume
that $g$ is smaller but not much smaller that
$g \approx 2(\frac{\L}{\ln \L})^2$, so that we can neglect the term on the
rhs of \eq{2d-critical}.
Then the critical line should be given 
approximately by
\be
m^2 = - \frac{3}{2}\;\sqrt{\frac{g }{\pi}}\;\L,
\label{2d-critical-approx}
\ee
or a slightly modified formula using the fuzzy torus regularization
(due to the different propagator).
Unfortunately we cannot compare this with the numerical results of 
\cite{bieten}, who consider a different scaling. 
However we can compare it to some extent
with  the numerical results of
\cite{martin} on the fuzzy sphere, and find reasonable agreement.
This will be done in the following section.

If $g$ scales like 
\eq{scaling-crit-2d}, one can use a simpler argument
due to \cite{martin} 
neglecting the kinetic term altogether, and replace the action
by the pure potential model $Tr V(\phi)$; hence this phase was denoted as 
``matrix phase'' in \cite{martin}.
Note that indeed the scaling 
\eq{scaling-crit-2d} is appropriate for the matrix model 
\eq{phi4-renorm} without the kinetic term.
This amounts to identifying 
\be
m'^2 = 2\pi\theta \frac{m^2}N, \quad g'  = 2\pi\theta\frac{g}N,
\label{pure-matrix}
\ee
which would predict a phase transition at 
$(2\pi\frac{m^2 \theta}{N})^2 = 8 \pi \frac{g \theta}{ N}$, i.e.
\be
m^2 = - \frac 1{\pi}\sqrt{2 g}\; \L.
\ee

\subsubsection{The fuzzy sphere}

For the case of the fuzzy sphere, we consider the action 
\be
S = \frac{4\pi R^2}{N} Tr(\phi \D \phi + r \phi^2 + \l \phi^4)
\ee
using the (redundant) parameters $r,\l, R$ following \cite{martin}.
The eigenvalues of $\D = \frac{{\cal L}^2}{R^2}$ are $l(l+1)/R^2
= p^2$, with cutoff $\L = p_{max} = N/R$. 
Comparing with \eq{phi4-renorm}, the above parameters
are related to the coupling constants $m$ and $g$ 
in \eq{phi4-renorm} via 
\be
\theta = \frac{2 R^2}{N}, \qquad 2r =  m^2, \qquad 4\l =  g.
\label{martin-ident}
\ee 
It turns out 
that $\a_0^2$ for the fuzzy sphere \eq{alpha-lattice-FS}
agrees precisely with the
result for $\R^2_\theta$, 
\be
\a_0^2(m_R) = \frac{1}{\pi}\ln(1+\frac{\L^2}{m_R^2}).
\ee
We can therefore apply \eq{2d-critical-approx}, assuming that 
$\l \ll (\frac{\L}{\ln \L})^2$; 
in particular we can fix $\l =1$ as in  \cite{martin}. 
Then the critical line is
\be
\frac{r}{N} = - \frac{3 }{2\sqrt{\pi}}\;\frac 1{R} \;
   \approx \; - 0.846\; \frac 1{R}.
\label{2d-critical-approx-fuzzy}
\ee
Note that $R$ is a free parameter corresponding to $\theta$, and we
should take at least $R = O(\sqrt{N})$. 

Let us compare his with Martin's disordered-matrix transition \cite{martin}:
he finds numerically a phase transition for (see eq. (46) in \cite{martin})
\be
\frac rN \approx -0.56\; \frac 1R
\label{martin-critical}
\ee
for large $R$.
This is reasonably close to our analytical result.
Recall that we cannot expect our prediction \eq{2d-critical-approx-fuzzy}
 to be exact since the eigenvalue
distribution is quite far from Wigner's law, and moreover
the arguments in 
Section 2 are weaker in 2 dimensions compared to 4 dimensions since
the crucial divergences are only logarithmic.
Furthermore $g$ strongly violates 
$g \sim 2(\frac{\L}{\ln \L})^2$ and therefore $m_R \to 0$, 
so that the replacement of the
kinetic term with \eq{eff-act-eig} is not justified here. 
The pure matrix model discussed above \eq{pure-matrix}
would give
\be
\frac{r}{N}  = - \frac{\sqrt{2}}{\pi}\; \frac 1R \;
  \approx \; - 0.45 \; \frac 1R.
\ee
Hence the numerical results of \cite{martin}
(taken for $N \approx 30$) appear to be in between the pure potential
model and our approach, and our treatment apparently overestimates the
kinetic term. This
is not too surprising in 2 dimensions in view of the above remarks.

\section{Discussion and outlook}
\label{sec:discussion}

We presented a simple non-perturbative approach to scalar field
theory on Euclidean non-commutative 
spaces, based on certain matrix regularizations of 
$\R^{2n}_\theta$. 
Starting from a representation of the field $\phi = U (\phi_i) U^{-1}$
in terms of eigenvalues $\phi_i$ and ``angles'' $U$, we observe
that the different behavior of planar and
non-planar diagrams due to UV/IR mixing implies a particular eigenvalue
density distribution, which can be reproduced by a simple
matrix model.
This is shown starting with the case of free 
fields, which can be described by a Gaussian matrix model.
Interactions of the form $Tr V(\phi)$ can then be included very
easily, and modify the eigenvalue distribution. 
This leads to a picture where the basic properties of the QFT such as 
correlation length and renormalization are related to its eigenvalue
density $\rho(s)$, through a ``constrained field theory''
with compact configuration space $\cO(\phi)$.
It also makes the existence of well-behaved scaling limits of NCFT
i.e. renormalizability very plausible.
We conclude that the eigenvalue sector of
non-commutative scalar field theories is  goverened by
an ordinary matrix model, which
provides a simple and intuitive
window into the non-perturbative domain.

For weak coupling, this approach provides a new way of computing
the renormalization of the potential; 
in particular, using a very simple approximation we
found an expression for the mass renormalization, which coincides with
the conventional one-loop calculation.
Furthermore, we found a phase transition at strong coupling
in the $\phi^4$ model in both 2 and 4 
dimensions, which is identified with the striped or matrix phase of 
\cite{gubser,martin}. This is particularly interesting in the
4-dimensional case since the critical line then
terminate at a non-trivial point with $g_c \neq 0$, which is interpreted as an
interacting RG fixed-point. 
The existence of this fixed point can be understood simply by 
noting that the critical line is characterized by a specific
eigenvalue distribution which is different from Wigner's law, which
holds for $g=0$. Therefore the critical line cannot end at $g=0$. 
All this 
suggests that such NC field theories may in fact be more acccesible to
analytical tools than their commutative counterparts. 

Perhaps the main shortcoming of this approach is the lack of a precise
relation between the eigenvalue distribution and the relevant
physical parameters, such as correlation
length resp. mass and coupling strength. 
It is quite clear that the leading parameter is the size $\a$ of
the maximal eigenvalue, which has been used in this paper 
to extract physical information. To go beyond this
approximation may be difficult, and may require e.g. perturbative methods.

There are many other questions and gaps which should be addressed
in future work. For example, the assumption that $\theta_{ij}$ is 
non-degenerate (or special) is quite clearly not essential, and a
similar approach should also work in odd dimension. Furthermore, a more
elaborate analysis of the renormalization in the weakly-coupled regime
should be attempted. Another interesting question concerns the relation
of this approach with the results of \cite{wulki} on a modified
$\phi^4$ model; to address this
issue, the above analysis should be repeated 
with a suitably modified propagator. This will be done elsewhere.

Perhaps the most interesting perspective is 
the possibility that careful estimates of the contribution of the kinetic
term on the orbits $\cO(\phi)$
(for eigenvalue distributions different from Wigner's law)
should allow to rigorously justify the above picture also in
the non-perturbative domain, and in particular the existence of
the critical point. One may in particular try to establish
renormalizability in this way.
This should be facilitated by the fact that
$\cO(\phi)$ are compact spaces.

Finally, it would of course be extremely
interesting to compare all this with numerical results in 4
dimensions, which are not available at this time.

\subsection*{Acknowledgements} 

I would like to thank H. Grosse, I. Sachs, and R. Wulkenhaar
for very useful discussions, and C-S. Chu for reading the manuscript.

\section*{Appendix A: Regularizations of $\R^{2n}_\theta$}
\addcontentsline{toc}{section}{Appendix A: Regularizations of $\R^{2n}_\theta$}

\subsection*{The fuzzy torus $T^2_N$ and $T^2_N \times T^2_N$}
\addcontentsline{toc}{subsection}{The fuzzy torus $T^2_N$ and $T^2_N \times T^2_N$}
\label{subsec:fuzzytorus}

A particularly simple regularization of $\R^{d}_\theta$ which works in
any (even) dimensions was proposed in
\cite{ambjorn}, which we review for convenience.
Consider a (toroidal) 
lattice with lattice constant $a$ and $N$ sites in each dimension. 
We denote its size with 
\be
L = Na.
\ee
Since we are on a torus, one
should not use the unbounded operators $x_{j}$. Instead consider
the unitary generators
\begin{equation}  
Z_{j} := e^{\frac{2\pi}{L} i x_j}\,, \qquad  Z_{i}^N =1.
\end{equation}
The commutation relations $[x_i,x_j] = i \Theta_{ij}$ then become
\be
Z_i Z_j  = \exp\(-\frac{4\pi^2}{L^2} i \Theta_{ij}\)  Z_j Z_i.
\label{ZZ-CR}
\ee
Rather than going through the most general case, 
we simply consider 
\be
\Theta_{ij} = \theta\; Q_{ij}
\ee
and work out the 4-dimensional case, where
\be
Q_{ij} = \begin{pmatrix}0 & 1 & 0  & 0    \cr
-1  & 0 & 0 & 0  \cr
0  & 0 & 0 & 1  \cr
0 & 0 & -1 & 0 \cr
\end{pmatrix}.
\ee
The generalization to any even dimension is obvious.
Periodicity  implies a quantization of $a$ resp. $\theta$ as
\be
\frac{N a^2}{\pi} = \theta,
\label{theta-lattice}
\ee
assuming that $N$ is odd.
The physical momentum is
\be
k_i = \frac{2\pi n_i}{aN}\; \in (-\frac{\pi}a, \frac{\pi}a )
 = (-\L,\L),
\ee
with UV-cutoff at
\be
\L = \frac{\pi}a  = \sqrt{\frac{\pi N}{\theta}}.
\label{NC-lattice-cutoff}
\ee
This is in qualitative agreement with the scaling \eq{cutoff-CPn}
obtained using fuzzy $\C P^n$.

In order to write down the action for a scalar field, we also need 
partial derivatives or shift operators,
\be
D_j := e^{a \partial_j}, \qquad 
 D_{j} Z_{i} D_{j}^{\dagger} 
   = e^{2\pi i \delta_{ij} /N} Z_{i} \,.
 \label{con2}
\ee
In our case, they can be  realized as 
\be
D_1 =  (Z_2^\dagger)^{(N+1)/2}, \quad D_2 =  (Z_1)^{(N+1)/2}
\ee
etc.
A solution of \eq{ZZ-CR} and $Z_i^N=1$ in 2 dimensions
is given by the unitary ``clock and shift'' operators (recall that $N$
is odd)
\begin{equation}
Z_{1} = \begin{pmatrix}0 & 1 &   &      &   &   \cr
  & 0 & 1 &      &   &   \cr
& & . & .  & & \cr
& & & . & . & & \cr
& & &  & 0 & 1 \cr
1 &&&&& 0 \end{pmatrix}, 
\qquad 
Z_{2} = \begin{pmatrix}1 &   &      &   &   &   \cr
  & e^{4\pi i/N} &   &   &      &   \cr
  &   & e^{2(4\pi i/N)} &      &   &   \cr
  &   &    & . &   &   \cr
  &   &    &   & . &   \cr
  &   &    &   &   & \end{pmatrix},
\end{equation}
which extends to arbitrary even dimensions by taking tensor products.
Hence the field $\phi$ is a hermitean $N^{d/2} \times N^{d/2}$ matrix.
The integral is again defined as 
\be
\int f = (2\pi\theta)^{d/2} Tr (f).
\ee
One can then define the ``plane waves'' 
\be
\phi_{\vec n} = \frac 1{N^{d/4}} \prod_{i=1}^d (Z_i)^{n_i}\; (\prod_{j<i}
e^{2\pi i  Q_{ij}n_i n_j/N})
\ee
which satisfy
\be
Tr(\phi_{\vec n}^\dagger \phi_{\vec n'}) = \d_{\vec n \vec n'}, \qquad
\phi_{\vec n}^{\dagger} = \phi_{-\vec n}
\ee
for $n_i \in [{-(N-1)/2,(N+1)/2}]$. They form a basis of the ``space of
functions'' $Mat(N^{d/2},\C)$.
Using
\be
D_{i} \phi_{\vec n} D_{i}^{\dagger} 
 = \exp(2\pi i  n_i/N)\; \phi_{\vec n}
\ee
one can write down the discretized lattice-version of \eq{Rd-action-star},
\bea
S[\phi ] &=&  (2\pi \theta)^{d/2}\; \Tr\left[\frac{1}{a^2}
\sum_{j=1}^{d}(\phi^2 - D_{j} \phi D_{j}^{\dagger}\phi)  +
  \frac{m^{2}}{2} \phi^{2} + \frac{g}{4} \phi^{4} \right].
\label{matact}
\eea
For hermitean 
$\phi = \sum p_{\vec n} \phi_{\vec n}$ with 
$p_{\vec n} = p_{-\vec n}^\dagger$, the kinetic term
becomes
\bea
\frac{1}{a^2} Tr \sum_i \left(D_{i} \phi D_{i}^{\dagger} - \phi \right)^{2} 
&=& \frac{2}{a^2} \sum_k |p_k|^2 \sum_j(1-\cos( k_j a)) \nn\\
&=& \sum_k |p_k|^2 (\sum_{j} k_j^2 + O(a^2 k^4)).
\label{kinetic-plane}
\eea
The propagator is therefore
\be
\langle p_{\vec k} p_{\vec k'}^\dagger \rangle 
= \d_{{\vec k} {\vec k'}}\; \frac 1{\frac{2}{a^2}\sum_j(1-\cos(k_j a)) + m^2},
\label{lattice-propagator}
\ee
and the phase factor for the nonplanar diagrams is obtained from
\be
\phi_{\vec k} \phi_{\vec k'} = \exp( -i \theta \sum_{i<j} k_i Q_{ij} k'_j)\;
\phi_{\vec k'} \phi_{\vec k}.
\ee

\subsubsection*{$\a_0^2$ on the fuzzy torus}
\label{subsec:torus-a2}

Due to the different  behavior of the propagator for large momenta,
$\a^2_0$ on the fuzzy tori will be somewhat different from 
regularizations using a sharp cutoff, such as on fuzzy $\C P^n$.
In 2 dimensions, 
we should compute more carefully
\bea
\langle \int d^2 x \phi^{2}(x) \rangle 
&=& V\;\int_{-\pi/a}^{\pi/a} \frac{d^2 p}{(2\pi)^2}\; \frac {a^2}{2\sum_i (1-cos(p_i
  a)) +m^2 a^2} \nn\\
&=&  V\;\int_0^{\pi} 
\frac{d^2 r}{(2\pi)^d}\; \frac {1}{\sum_i (1-cos(r_i)) +m^2 a^2/2} \nn\\
\label{phi2-expect-lattice-2d}
\eea
and in 4 dimensions
\bea
\langle \int d^4 x \phi^{2}(x) \rangle 
&=& V\;\int_{-\pi/a}^{\pi/a} \frac{d^4 p}{(2\pi)^4}\; \frac {a^2}{2\sum_i (1-cos(p_i
  a)) +m^2 a^2} \nn\\
&=& \frac{V}{\pi^2}\;\L^2\; \int_0^{\pi} 
\frac{d^4 r}{(2\pi)^4}\; \frac {1}{\sum_i (1-cos(r_i)) +m^2 a^2/2} \nn\\
\label{phi2-expect-lattice-4d}
\eea
for the above regularization.
In particular, for $m=0$ one has
\be
\a_0^2(m) \leq
\a_0^2(m=0) = \frac{4}{\pi^2}\;\L^2\; 
\int_0^{\pi} \frac{d^4 r}{(2\pi)^4}\; \frac {1}{\sum_i (1-cos(r_i))} 
= \frac{ 0.31}{4\pi^2}\;\L^2
\label{alpha-lattice-4d}
\ee
numerically. This is needed e.g. to determine the critical coupling
$g_c$ for the $\phi^4$ model in 4 dimensions.

\subsection*{The fuzzy sphere}
\addcontentsline{toc}{subsection}{The fuzzy sphere}
\label{sec:fuzzysphere}

The algebra $S_N^2$ of
functions on the fuzzy sphere \cite{madore}
is the finite algebra 
generated by Hermitian operators
$x_i= (x_1, x_2, x_3)$ satisfying the defining relations
\bea
[x_i, x_j] = i \L_N \e_{ijk} x_k, \label{def1}\\
x_1^2 + x_2^2 +x_3^2  = R^2 \label{def2}
\eea
where $R$ is an arbitrary radius.
The noncommutativity parameter $\L_N$ is of dimension length, and  
is quantized by
\be\label{def3}
\frac {R}{\L_N} = \sqrt{\frac{N^2-1}{4}} \; ,
\quad \mbox{$N = 1,2,\cdots$ }
\ee
This can be easily understood: \eq{def1} is
simply the Lie algebra $su(2)$, whose irreducible representation
have dimension $N$. The Casimir
of the $N$-dimensional representation is quantized, and related to $R^2$ by
\eq{def2} and \eq{def3}.
Thus the fuzzy sphere is characterized by its radius $R$ and the
``noncommutativity parameters'' $N$ or $\L_N$.
The algebra of ``functions'' $S_N^2$ is simply the
algebra $Mat(N)$ of $N \times N$ matrices.
It is covariant under the adjoint action of $SU(2)$, under which it
decomposes into the irreducible representations with dimensions
$(1) \oplus (3) \oplus (5) \oplus ... \oplus (2N-1)$.
The integral of a function $f \in S_N^2$ over the fuzzy sphere is 
\be
\int \phi(x)  = \frac{4 \pi R^2}{N} Tr[\phi(x)],
\ee
which agrees with the integral on $S^2$ in the large $N$ limit
and is invariant under  the  $SU(2)$ rotations.
The dimensionless coordinates
$\l_i = x_i/\L_N$
generate the rotation operators $J_i$:
\beq
J_i f = [\l_i, f].
\label{J-def}
\eeq
One can now easily write down actions for scalar fields, such as 
\be
S  = \frac{4\pi R^2}N Tr\(\frac 12 \phi \D \phi   
   + \frac 12 m^2 \phi^2 + \frac 14 g \phi^4\) 
 = 2\pi \theta\; Tr\(\frac 12 \phi \D\phi   
   + \frac 12 m^2 \phi^2 + \frac 14 g \phi^4 \)  
\label{action-S2-dimless}
\ee
were $\phi$ is a Hermitian matrix, and
$\D \phi =  \frac 1{R^2} J_i J_i \phi$  is the Laplace operator.
The last form is obtained by defining $\theta$ through
\be
R^2 = \frac{N \theta}2.
\ee
There are 2 obvious  $N \to \infty$  limits: 1) the conventional,
commutative $S^2$ limit keeping  $R$ fixed,
and 2)  the limit of the  NC plane $\R^2_\theta$, which can be obtained
by keeping $\theta$ constant: then the tangential 
coordinates $x_{1,2}$ satisfy at the north pole the commutation relations 
\be
[x_1, x_2] = i \frac{2R}N x_3 = i \frac{2R}N \sqrt{R^2-x_1^2-x_2^2} 
\approx i \theta .
\ee
Note that $\theta^{-1}$ now determines the basic (NC) scale of the NC field
theory, replacing the radius $R$. 

If one considers non-planar loops, the oscillatory
behavior due to the $6j$ symbols
sets in for angular momenta 
$l^2 \approx N$ and was studied in \cite{fuzzyloop}.
Note that this corresponds to
$p^2 = O(\theta^{-1})$
in the NC case, but to $p \to\infty$ in the commutative case. Therefore 
the non-oscillatory domain is  divergent in the
commutative limit and prevents a
matrix behavior. 
This ``low-energy'' sector is suppressed in the NC
case if $\theta$ is kept finite.

\subsubsection*{$\a_0^2$ for the fuzzy sphere}
\label{subsec:FS-a2}

The arguments of Section \ref{subsec:free distri} 
for the eigenvalue distribution go 
through here as well provided the oscillating behavior of the
non-planar diagrams is strong enough. 
Even though there are nontrivial effects even for finite $R$ resp.
$\theta \propto 1/N$ \cite{fuzzyloop}, they are sufficiently strong 
for our purpose only if $\theta$ is kept finite.
To find the appropriate $\a_0(m)$, consider for $g=0$
\bea
\langle \int_{S^2}\phi^{2}\rangle 
&=& \sum_{l,m}\int_{S^2} \frac{Y^{lm} \cdot Y^{l-m}}{l(l+1)/R^2 + m^2} 
 = \sum_{l=1}^N \frac{2l+1}{l(l+1)/R^2 + m^2} \nn\\
&=& R \sum_{l=1}^N \frac{(2l+1)/R}{l(l+1)/R^2 + m^2} 
= 2 R^2 \int_0^{N/R} dx \frac{x}{x^2 + m^2} \nn\\
&=&  R^2  \ln(1+\frac{N^2}{m^2 R^2}) 
= \frac{V}{4 \pi} \ln(1+\frac{\L^2}{m^2}) 
\label{alpha-lattice-FS}
\eea
in the large $N$ limit,
using $V = 4 \pi R^2$ and $x = \frac{l+1/2}{R}$ and 
$\L = \frac NR$. This agrees precisely with \eq{phi2-expect-2d}, so that the 
same $\a_0(m)$ as in \eq{alpha} can be used.

\subsection*{Fuzzy $\C P^n$}
\addcontentsline{toc}{subsection}{Fuzzy $\C P^n$}
\label{sec:fuzzyCPn}

The construction of fuzzy $\C P^n$ \cite{stroh,cpn-bala,wataCPN,CP2-gauge} 
is analogous to the case of 
fuzzy $S^2 \cong \C P^1$. 
Since $\C P^n$ is an adjoint orbit of $SU(n+1)$, 
it is a compact symplectic space and can be quantized in terms 
of finite matrix algebras $Hom(V_N)$ where
$V_N$ are suitable \reps of $su(n+1)$. To identify the correct \reps
$V_N$ of $su(n+1)$, we must match the space of harmonics on classical 
$\C P^n$
\beq
\cC^{\infty}(\C P^n) = \mathop{\oplus}_{p=0}^\infty V_{(p,0,...,0,p)}.
\label{CPn-harmonics}
\eeq
with the decomposition of 
$Hom(V_N)$.
It is easy to show that indeed
\beq
Hom(V_N) = V_N \tens V_N^*\cong \mathop{\oplus}_{p=0}^{N}\; V_{(p,0,....,0,p)}
\label{decomp_CPn}
\eeq
for
$$
V_N := V_{(N,0,...,0)}.
$$
Here $V_{(l_1,...,l_n)}$ denotes the highest weight irrep of $su(n+1)$
with highest weight $l_1 \L_1 + ... + l_n \L_n$ where $\L_k$ are the 
fundamental weights. 
One can therefore
define the algebra of functions on the fuzzy projective space by
\beq
\C P^n_N := Hom_\C(V_N) = Mat(\cN,\C)
\eeq
with
\be
\cN = \frac{(N+n)!}{N! n!} \approx \frac{N^n}{n!}.
\label{n-cN-CPn}
\ee
The functions on fuzzy $\C P^n$ have a UV cutoff given by $N$. 
Scalar fields on $\C P^n_N$
are elements in $Hom_\C(V_N)$, and
the integral is given by the suitably normalized trace over $V_N$.
The coordinate functions $x_a$ for $a=1,...,n^2+2n$
on fuzzy $\C P^n$ are given by suitably rescaled generators of
$su(n+1)$ acting on $V_N$.
One finds \cite{wataCPN}
\beqa
[x_a, x_b] &=&  i \Lambda_N f_{abc}\; x^c, \qquad
g^{ab} x_a x_b    = R^2, \label{defz2-n} \\
d^{ab}_c x_a x_b 
&=& (n-1)(\frac{N}{n+1} +\frac 12) \Lambda_N\;x_c. \label{defz3-n}
\eeqa
for 
\beq
\Lambda_N = \frac R{\sqrt{\frac{n}{2(n+1)} N^2 +\frac n2 N}}.
\eeq
For large $N$, this reduces to the defining relations 
of $\C P^n \subset \R^{n^2+2n}$.
On the other hand, scaling the radius as 
\be
R^2 = N \theta \frac{n}{n+1}
\label{R-N-CPn}
\ee
near a given point (the ``north pole'') of $\C P^2_N$ gives
$\R^{2n}_\theta$ with $U(n)$ invariant $\theta_{ij}$,
similarly as for the fuzzy sphere. We refer to \cite{CP2-gauge} 
for further details.

The Laplacian on fuzzy $\C P^n$ is proportional to the quadratic Casimir of 
$su(n+1)$ acting on the functions,
\beq
\Delta(\phi) = \frac {c}{R^2}\; J_a J_a \phi,
\eeq
where $J_a$ generates the $SU(n+1)$ rotations and
$c = \frac{2n}{n+1}$.
It has eigenvalues
\be
\D f_k(x) 
= c \frac {k (k+n)}{R^2} \; f_k(x)
\ee
for $f_k(x) \in V_{(k,0,...,0,k)}$
according to the decomposition \eq{decomp_CPn}.
The multiplicity for given $k$ is
\be
dim(V_{(k,0,...,0,k)}) = \frac{(k+n-1)!^2}{k!^2 (n-1)!^2} \frac{2k+n}{n}
 \approx \frac{2}{(n-1)!^2 n}\; k^{2n-1}
\ee 
for $k \gg n$. 
To find the appropriate $\a_0(m)$, consider 
\bea
\langle \int_{\C P^n}\phi^{2}\rangle 
&=& \sum_{k,m}\int_{\C P^n} \frac{Y^{km} \cdot Y^{k-m}}{c k(k+n)/R^2 + m^2} 
 = \frac{2}{(n-1)!^2 n}\;\sum_{k=1}^N \frac{ k^{2n-1}}{c k(k+n)/R^2 +  m^2}  
         \nn\\
&=& \frac{2 (R/\sqrt{c})^{2n-1} }{(n-1)!^2 n}\;\sum_{k=1}^N 
  \frac{ (k\sqrt{c}/R)^{2n-1}}{c k(k+n)/R^2 + m^2} 
= \frac{2(R/\sqrt{c})^{2n}}{(n-1)! n!}\; 
   \int_0^{\L} dx \frac{x^{2n-1}}{x^2 + m^2}   \nn\\
&=& \frac{V}{2^{2n-1}\pi^n (n-1)!}
  \int_0^{\L} dx \frac{x^{2n-1}}{x^2 + m^2}  
\label{CPn-comput}
\eea
in the large $N$ limit, where we denote the basis again with $Y^{km}$ and
used $V = Vol(\C P^n) = \(\frac{2(n+1)}{n}\)^n \frac{\pi^n}{n!}\;R^{2n}$ 
and $x = \sqrt{c} \frac{k}{R}$ and 
\be
\L = \sqrt{c}\frac NR = \sqrt{\frac{2n}{n+1}}\;\frac NR
 = \sqrt{\frac{2N}\Theta}.
\label{cutoff-CPn}
\ee
This agrees with precisely with \eq{phi2-expect-2d} and generalizes
the results for the fuzzy sphere. Furthermore, 
putting \eq{cutoff-CPn} and \eq{R-N-CPn} 
together gives
\be
\int_{\C P^n_N} := \frac{V}{\cN} Tr(.) \to (2\pi \theta)^n Tr(.)
\ee
in the above scaling limit.
Therefore this rescaled 
fuzzy $\C P^n$ is a perfect regularization of 
$\R^{2n}_\theta$. It has a sharp mumentum cutoff at $\L$, and the
same $\a_0(m)$ as in \eq{alpha} can be used.

\section*{Appendix B: Justification of \eq{RNP-lemma}}
\addcontentsline{toc}{section}{Appendix B: Justification of \eq{RNP-lemma}}
\label{sec:nonplanar-appendix}

Consider first $d \geq 3$. Then
\be
R_{NP} :=\frac{\langle  \frac 1V\int d^d x \phi(x)^{2n} \rangle_{Non-Planar}}
{\langle  \frac 1V\int d^d x \phi(x)^{2} \rangle^n}
 \approx \frac 1c\;\sum \int_0^1 \frac{d^dk'_{1}}{(k'_1)^{2}} ... \frac{d^d k'_n}{(k'_n)^{2}}
e^{i \L^2\sum k'_i \Theta k'_j},
\label{RNP-append}
\ee
and we assume that $\L^2 \gg m_\theta^2, \; \L^2 \gg m^2$.
Here the integration domains have been rescaled to be unit balls in
momentum space for all diagrams in the numerator and in the
denominator; then the denominator can be estimated
by the planar contribution which
gives a finite contribution $c$ (after the rescaling), which will be omitted. 
To proceed,
let  $\vec v \propto (k_1. ..., k_{nd}) \in
\R^{nd}$ denote a unit vector in $\R^{nd}$
with norm $\|\vec v\| =1$, and consider generalized spherical coordinates
such that
$d^d k_i ... d^d k_n = d\Omega(\vec v)\; r^{nd-1} dr$ in $\R^{nd}$.
Then
\be
 \int_0^1 \frac{d^dk'_{1}}{(k'_1)^{2}} ... \frac{d^d k'_n}{(k'_n)^{2}}
e^{i \L^2\sum k'_i \Theta k'_j} 
 = \int d\Omega(\vec v) \int_0^{O(1)} dr \; 
r^{nd-1} \frac{e^{i \L^2 r^2 \vec v \Theta \vec v }}{(k'_1)^{2} ... (k'_n)^{2}}
\ee
in simplified notation.
The important point is that the radial integral $\int dr$ is
oscillatory,  with increasing frequency. For $d \geq 3$, 
the radial integral behaves as
\be
\int_0^{O(1)} dr \;r^{n(d-2)-1} e^{i \L^2 r^2 \vec v \Theta \vec v}
 = \int_0^{O(1)} du \;u^{n(d-2)/2-1} e^{i \L^2 u \vec v \Theta \vec v}
\label{radial-estimate}
\ee
where $r^2  =u$. Clearly the (alternating) contributions increase with
$u$. The integral is therefore estimated by the ``last oscillation'' which
is of order $min\{1,O(1/(\L^2 \vec v \Theta \vec v))\}$.
We can exclude the region $\{\vec v \Theta \vec v <  1/\L\}$, whose
volume goes like $O(1/\L)$ as $\L \to \infty$ 
(ignoring $\log \L$-corrections). 
Therefore 
\be
R_{NP} \leq O(1/\L )
\ee
ignoring possible $\log$-corrections\footnote{this is probably not the
best possible estimate}, which establishes our claim.

For $d=2$, we apply the same analysis to the numerator of
\eq{RNP-append}. The radial integral has again the form
\eq{radial-estimate}, but it is now dominated by small
$u$, i.e. the ``first half-oscillation''. We therefore
have to reintroduce the masses which provide an IR cutoff.
To estimate this, we go back to the original form
\bea
\langle  \frac 1V\int d^d x \phi(x)^{2n} \rangle_{Non-Planar} &=& 
\int_0^\L \frac{d^2k_{1}}{(k_1)^{2}+m^2} ... \frac{d^2 k_n}{(k_n)^{2}+m^2}
e^{i \sum k_i \Theta k_j} \nn\\
&\approx&
\int_{\sum k_i \Theta k_j <\pi} \frac{d^2k_{1}}{(k_1)^{2}+m^2}
... \frac{d^2 k_n}{(k_n)^{2}+m^2} \nn\\
 &\approx& f(\ln(\frac{m}{m_\theta})).
\eea
Indeed 
the result can only depend on the ratio $\frac{m}{m_\theta}$ where
\be
m_\theta^2 = \frac 1{\theta}
\ee
is the NC mass scale, and in 2 dimensions it will depend only logarithmically
on this ratio (for example, 
$\langle  \frac 1V\int d^d x \phi(x)^{4} \rangle_{Non-Planar} = O(\ln(\frac{m}{m_\theta})^2$).
On the other hand,
\be
\langle  \frac 1V\int d^2 x \phi(x)^{2n} \rangle_{Planar} 
 = O(\ln(\frac{\L}m)^n),
\ee
Hence for fixed $\frac{m}{m_\theta}$ and $\L \to\infty$, the
eigenvalue distribution is again given by the Gaussian matrix model
\eq{eff-act-eig}, however we expect the fluctuations to be larger in this case
than for $d=4$.

It is interesting to consider also 
the commutative scaling limit of the 
fuzzy sphere, in view of the ``non-commutative anomaly'' found in
\cite{fuzzyloop}. In that case, we can set $\theta = 1/N R^2$ and therefore 
$m_\theta^2 = O(N) = O(\L)$. Then both planar and non-planar
contributions have the same logarithmic behavior, and  there is
no well-defined eigenvalue distribution in that case. This was to be
expected since we considered the commutative limit; however the 
``non-commutative anomaly'' indicates some NC behavior in
that case too, which is apparently too weak to 
induce a distinct eigenvalue distribution. 
The situation is different in 4 dimensions, and a well-defined
eigenvalue distribution may well exist in the commutative scaling limit.

\end{document}